\documentclass[journal=jpclcd,manuscript=letter]{achemso}
\usepackage[version=3]{mhchem} 
\usepackage{natbib}

\usepackage{amsfonts}
\usepackage{amssymb}
\usepackage{graphicx}
\usepackage{subfigure}
\usepackage{amsmath}
\usepackage[english]{babel}
\usepackage{color}
\usepackage{epstopdf}
\usepackage{footnote}

\author{Kaijun Shen}
\affiliation[First University]{School of Materials Science and Engineering, Nanyang Technological University, Singapore 639798, Singapore}
\author{Kewei Sun}
\affiliation[Second University]
{School of Science, Hangzhou Dianzi University, Hangzhou 310018, China}
\author{Maxim F. Gelin}
\affiliation[First University]
{School of Materials Science and Engineering, Nanyang Technological University, Singapore 639798, Singapore}
\alsoaffiliation[Second University]
{School of Science, Hangzhou Dianzi University, Hangzhou 310018, China}
\author{Yang Zhao}
\affiliation[First University]
{School of Materials Science and Engineering, Nanyang Technological University, Singapore 639798, Singapore}
\email{YZhao@ntu.edu.sg}

\title{Finite-Temperature Hole-Magnon Dynamics in an Antiferromagnet}

\begin{document}


\maketitle
\begin{tocentry}
\centering
\includegraphics[width = 5.08cm, height = 5.08cm]{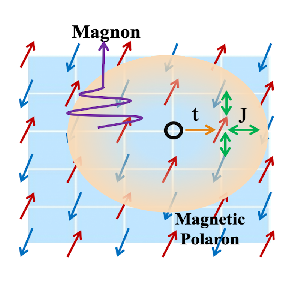}
\end{tocentry}

\begin{abstract}
Employing the numerically accurate multiple Davydov Ansatz in combination with the thermo-field dynamics approach, we delve into interplay of the finite-temperature dynamics of holes and magnons in an antiferromagnet, which allows for scrutinizing previous predictions from self-consistent Born approximation while offering, for the first time, accurate finite-temperature computation of detailed magnon dynamics as a response and a facilitator to the hole motion. The study also uncovers pronounced temperature dependence of the magnon and hole populations, pointing to the feasibility of potential thermal manipulation and control of hole dynamics.
Our methodology can be applied not only to the calculation of steady-state angular-resolved photoemission spectra,
but also to the simulation of femtosecond terahertz pump-probe and other nonlinear signals for the characterization of antiferromagnetic materials.
\end{abstract}



Deciphering the dynamics of charge carriers within quantum spin environments constitutes a pivotal aspect of contemporary condensed matter physics research~\cite{naga, schm, shra, kane, sach, trug, liu, dago, Wen, mish, mano, Pirro21}, with significant implications for high-temperature superconductivity (HTS) and exotic magnetic phenomena. The Fermi-Hubbard model, serving as an essential theoretical underpinning~\cite{boll, mazu, brown, chiu, vija, guard, koeps, gall}, in combination with high-resolution quantum gas microscopes~\cite{bakr, hall, yang},  facilitates detailed investigations into the structural and dynamical responses of lattice defects~\cite{chiu, geoffrey}.  A compelling characteristic of the Fermi-Hubbard model resides in its examination of hole dynamics and the associated emergence of magnetic polarons, both intricately linked to the fundamental physics of HTS~\cite{dago, Wen, schr}.
The multifaceted nature of these systems has ignited renewed interest~\cite{ande, brown, nich, chiu, koeps}. Noteworthy recent progress includes the utilization of a cold-atom simulator to observe a single hole's evolution in an antiferromagnet (AFM), unveiling rapid delocalization and magnon generation~\cite{geoffrey}.

The $t\text{-}J$ model, an adaptation of the Hubbard model, aid our grasp of spin-charge separation, string states, and {\emph{d}}-wave HTS~\cite{mano}. The exploration of the $t\text{-}J$ model navigates the challenging landscape of quantum many-body systems far from equilibrium, a complex class of problems persistently eluding comprehensive theoretical characterization. Intriguingly, the string excitations -- internal excitations of the hole polaron -- have been reported previously by high resolution angle-resolved photoelectron studies of cuprates~\cite{ronn, graf}. These string excitations originate in the hole's trajectory, yielding a trail of reversed spins that deviate from the local AFM backdrop.  The moving hole, hence, not merely introduces spin deviations, but it sequentially creates or annihilates a spatially continuous chain of such deviations, resulting in a linearly rising potential.

Nielsen \textit{et al}.~\cite{knak} employed the self-consistent Born approximation (SCBA) to compute the hole dynamics in an AFM, which was extended to handle two holes \cite{knak23} and bilayers \cite{knak23a}.
However, inter-site correlations are treated in SCBA in a mean-field manner, leading to a notable omission of the intricate dynamical details at ultra-short time scales and an insufficient representation of magnon-related phenomena. Furthermore, the SCBA is restricted to zero temperature, while the strive to comprehend microscopic origins of HTS ~\cite{dago, Wen, schr}, as well as the striking finding that a hole strongly coupled to spins creates and maintains quantum correlations even at infinite temperatures \cite{nagy}, motivate us to extend the description beyond $T=0$, notably taking into account that the available finite-temperature approaches are also based on the mean-field approximation \cite{finiteT91,finiteT93,lzyu}.

To address these challenges, we combine the multiple Davydov Ansatz (mDA) method~\cite{zhao1, zhao2} with the thermo-field dynamics (TFD) approach~\cite{Y,M1,Rams,Borrelli} for developing a numerically-accurate methodology capable of solving the $t\text{-}J$ model at finite temperatures. The
mDA approach is a well-established many-body wave-function variational method based on Gaussian states, which has been applied to a range of problems, from various Landau-Zener transitions~\cite{wang, zheng} and cavity polariton dynamics~\cite{sun1,sun2} to excitonic light harvesting~\cite{chen1,chen2} and ultrafast spectroscopy at conical intersections (CIs)~\cite{shen, sun5}.
Anfinite-temperature wave-function representation of quantum mechanics \cite{Y,M1,Rams}, TFD has been integrated with tensor train (TT) methods to become a powerful instrument for evaluating many-body quantum dynamics \cite{BG20,BG21,BG23,footnote,White05,Qin22,Sinha22}.
The combined mDA-TFD methodology has been applied to dynamics of Holstein polarons \cite{chen} and CI-assisted singlet fission systems \cite{LP}.
In this work, mDA-TFD is used to delineate a comprehensive portrait of the correlated hole-magnon dynamics in an AFM at finite temperatures.

We delve into the dynamics of a single hole within a fermionic gas comprised of two spin components arranged in a two-dimensional square lattice. As the interparticle repulsion significantly intensifies, these two spin components shape a quantum AFM. Under these conditions, the system's behavior can be captured by the $t\text{-}J$ model. Within a slave-fermion representation, the underlying problem is described by the Hamiltonian~\cite{kane,lzyu,zhao}
\begin{eqnarray}
\label{H}
\hat{H} &&= \sum_{\bold q} \omega_{\bold q} \hat{b}_{\bold q}^{\dagger} \hat{b}_{\bold q} + \frac {tz}{\sqrt{N}} \sum_{\bold {kq}} \hat{h}_{\bold {k-q}}^{\dagger} \hat{h}_{\bold k} [(u_\bold q \gamma_ {\bold {k-q}}+v_{\bold q} \gamma_{\bold k}) \hat{b}_{\bold q}^{\dagger}\nonumber \\
&&+(u_\bold q \gamma_{\bold k}+v_{\bold q} \gamma_{\bold {k-q}}) \hat{b}_{\bold {-q}}] ,
\end{eqnarray}
where $\hat{h}_{\bold q}^{\dagger}$ ($\hat{h}_{\bold q}$) is the hole creation (annihilation) operator, $\hat{b}_{\bold q}^{\dagger}$ ($\hat{b}_{\bold q}$) creates (destroys) a magnon with crystal momentum $\bold q$ and energy $\omega_{\bold q} = JzS \sqrt{1-\alpha^2 \gamma_{\bold q}^2}$, and
\begin{eqnarray}
u_{\bold q} = \sqrt{\frac{JzS+\omega_{\bold q}}{2 \omega_{\bold q}}},~~ v_{\bold q} = - {\rm sgn} (\gamma_{\bold q}) \sqrt{\frac{JzS-\omega_{\bold q}}{2 \omega_{\bold q}}}
\end{eqnarray}
are the coupling coefficients. The positive-valued $J$ represents the nearest-neighbor spin interactions and $t$ is the hopping strength. $z$ and $N$ denote the number of nearest neighbors (NNs) and the number of sites, respectively. We take the spin quantum number $S = 1/2$, $\alpha=1$ (Heisenberg limit indicating the isotropic spin-spin interactions) and the structure factor written in the for $\gamma_{\bold q} = [\cos(q_x)+\cos(q_y)]/2$ implies that the lattice constant is set to unity.
For Hamiltonian (\ref{H}), mDA of multiplicity $M$ can be written as follows
\begin{eqnarray}
\label{D2}
| D_2^{\rm M} (\tau) \rangle = \sum_{\scriptstyle1\le n\le N \atop\scriptstyle1\le m\le M }  A_{n m} (\tau) |n \rangle  {\rm e}^{\sum_{ \{ \bold q\} } (f_{m \{ \bold q\} }(\tau) b_{\{ \bold q\}}^\dagger - {\rm {H.c.}})}
| \bold 0 \rangle
\end{eqnarray}
where H.c.~stands for Hermitian conjugate, $| \bold 0 \rangle$ is the vacuum state for the magnons, $|n \rangle$ denotes the hole states, $\{ \bold q\}= \bold q \oplus \tilde{\bold q}$, $\tilde{\bold q}$ is the ``tilde" momentum  responsible for temperature effects in the TFD-mDA method, and $f_{m \{ \bold q\}}(t)$ is the magnon displacement with momentum $  \bold q$ and tilde momentum $\tilde{\bold q}$ in the $m$th coherent state (see Supporting Information for technical details).
In all calculations, an $8 \times 8$ lattice is adopted ($N=64$) and the NN number is fixed at $z=4$. In the $x-y$ reference frame with the origin in the center of the lattice, positions of all sites are determined by the radius-vector $\bold d$, and the hole dynamics is specified by the density matrix $\rho_h({\bold d}, \tau)$, which is normalized according to $\sum_{\bold d} \rho_h({\bold d}, \tau) = 1$.
All calculations are performed with the mDA multiplicity $M = 60$, which is required to tackle the doubling tilde modes from the finite-temperature  effect and sufficient for obtaining numerically converged results. The hole is initially placed in the center of the lattice ($\bold d=0$). Hence distances $d=0, \, 1, \, \sqrt{2}, \, 2$ correspond to  the initial hole site (IHS), NNs, next-nearest neighbors (NNNs), and second-nearest neighbors (SNNs), respectively.

\begin{figure}[htb]
\centering
\subfigure[]{
\includegraphics[scale=0.32,trim=50 50 50 0]{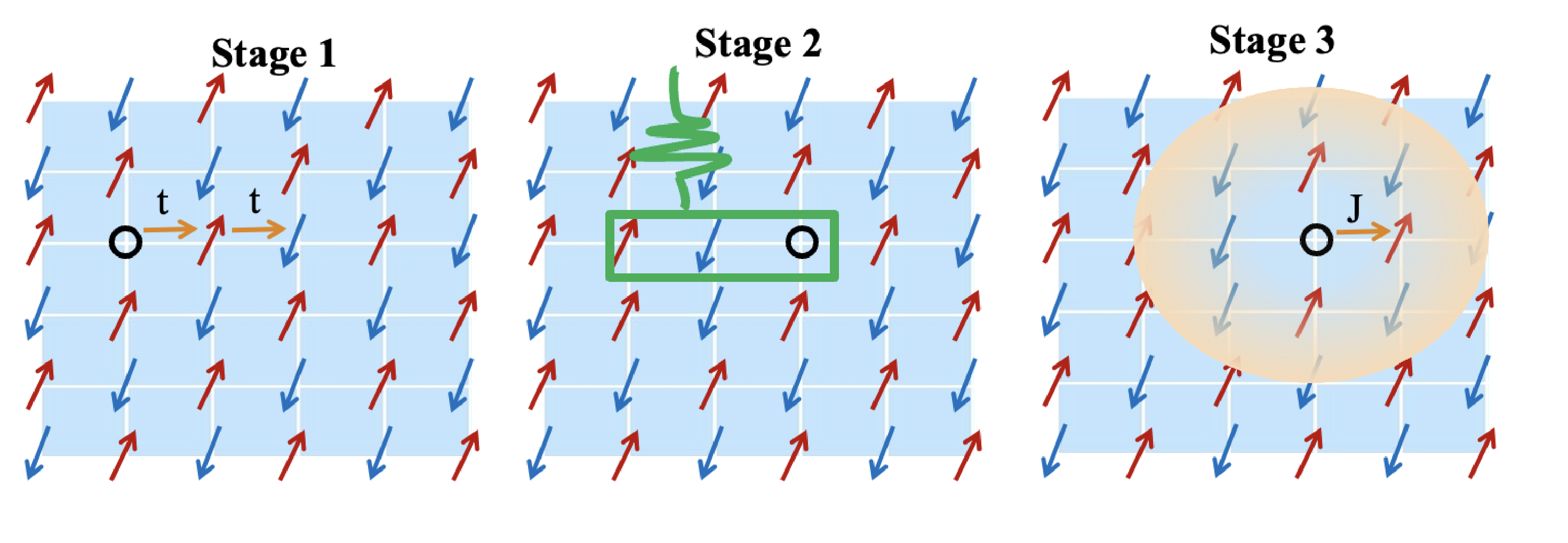} 
}\\
\includegraphics[scale=0.45,trim=0 0 0 10]{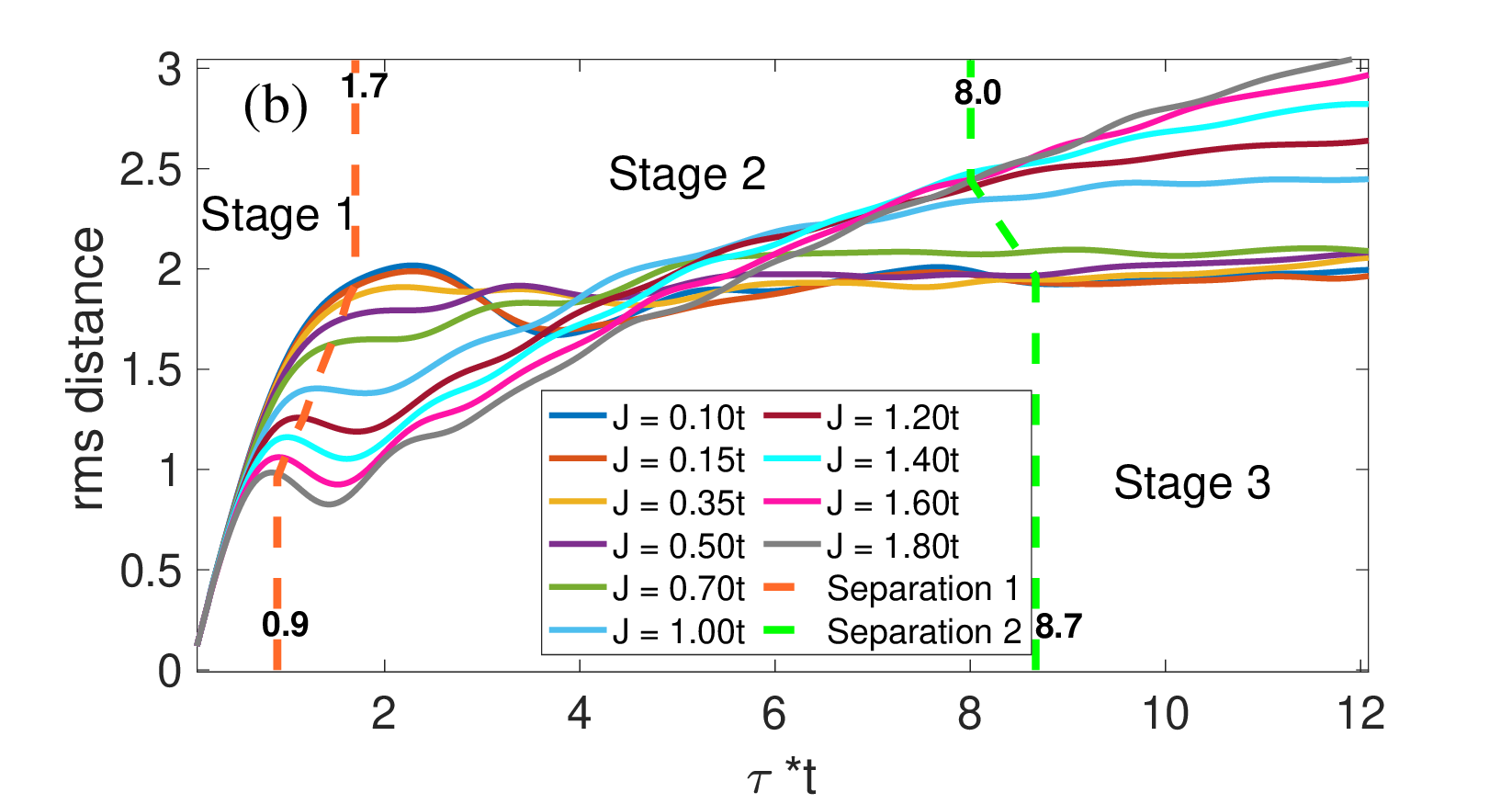}

\caption{Single-hole dynamics at zero temperature. (a) Sketch of three stages underpinning magnetic polarons emerging from localized defects: (i) kinematic expansion, (ii) polaron formation, and (iii) polaron drift. (b) Evolution of the rms distances $d_{\rm rms}(\tau)$ for spin interaction strengths $J$ indicated in the legend. }
\label{Fig1}
\end{figure}

We begin with the recapitulation of the hole dynamics at zero temperature. According to the experimental findings~\cite{geoffrey} and SCBA simulations~\cite{knak}, one can delineate three stages of the hole transport which are sketched in Fig.~\ref{Fig1}(a):  (i) initial kinematic expansion, (ii) emergence of strings and magnetic polarons, and (iii) ballistic low-speed polaron motion. Our mDA-TFD simulations confirm the existence of these three stages. This is illustrated by Fig.~\ref{Fig1}(b), which displays evolution of the root-mean-square (rms) distances $d_{\rm rms}(\tau) = \sqrt{\sum_{\bold d} \rho_h({\bold d}, \tau) d^2}$ for a range of spin interaction strengths $J$.
 In agreement with Ref.~\cite{knak}, the slope of $d_{\rm rms}(\tau)$ during the ballistic stage (i) deviates from the ideal value of $2t$ owing to interactions with spins. The timescale of this stage, $\tau_{(i)}$, is  proportional to $1/t$ in the strong-coupling regime ($J\ll t$), while $\tau_{(i)} \sim 1/J$ in the weak-coupling regime ($J\gg t$). For $J/t = 0.1$ and $J/t = 1.8$, for example, the initial ballistic motion ceases at $\tau  \approx  1.7/t$ and $\tau \approx 0.9/t$, respectively.
Stage (ii) is governed by the interference of strings (which correspond to higher-energy states in the hole spectral density) and polarons (which correspond to lower-energy states in the hole spectral density) \cite{mano,Diamantis21}.
The stage (ii) terminates at $\tau  \approx  8.7/t$ and $\tau  \approx  8.0/t$ for $J/t = 0.1$ and $J/t = 1.8$, respectively, where the corresponding transition time scales as $J/t$ for weak interactions and $(J/t)^{\frac {-2}{3}}$ for strong couplings~\cite{knak}.
This delineates the dynamic transition culminating in the eventual propagation of magnons during stage (iii) characterized by the time interval $\tau_{(iii)}$.

The values of $\tau_{(i)}$ and $\tau_{(iii)}$ are determined by the competition between two processes.
When the hole travels a displacement that spans an integral number $l$ of lattice constants, the spins subsequently arrange into a pattern incongruous with the Neel state. This deviation from the AFM norm elevates system's exchange energy by a magnitude on the order of $lJ$. The energetic cost of this disruption effectively inhibits the hole from vacating its initial position, where the Neel order remains undisturbed. Consequently, this energy landscape results in the autolocalization~\cite{lzyu} of the hole at a longer time.
Basically, a hole placed in a lattice with diminished spin-spin coupling $J$ is likely to traverse further from its primary position prior to the influence exerted by the inherent spin order of the quantum magnet. Nevertheless, a smaller $J$ concurrently signifies a more profound dressing of the hole by spin waves in its final polaron state.
For   small $J$, the hole dressing wins over the decreased $\sim lJ$ energy penalty in the beginning of stage (ii), preventing the hole from leaving its initial position and pushing it back. This dynamic event is manifested through a pronounced partial recurrence of $d_{\rm rms}(\tau)$ in Fig.~\ref{Fig1} (b) which is not grasped by the SCBA method~\cite{knak}. This discrepancy can be attributed to the limitations of SCBA in accurately addressing collective effects and long-range order in strongly correlated systems. A larger $J$ diminishes the dressing effect, causing quenched oscillations in $d_{\rm rms}(\tau)$, the amplitudes of which decrease with $J$. This agrees with predictions of Ref.~\cite{knak}, though SCBA underestimates the oscillation amplitudes in $d_{\rm rms}(\tau)$ in the intermediate regime $t \sim J$ (cf. Ref. \cite{Bohrdt20}).
Overall, the magnon dressing of the hole significantly decelerates its ballistic expansion in stage (iii) in the strong-coupling regime, leading to the quasi-trapping of the hole. In the weak-coupling regime, the hole moves ballistically in stage (iii)) ($d_{\rm rms}(\tau)\sim \tau$), though much slower than in stage (i).

\begin{figure}[htb]
	\centering
	\includegraphics[scale=0.45]{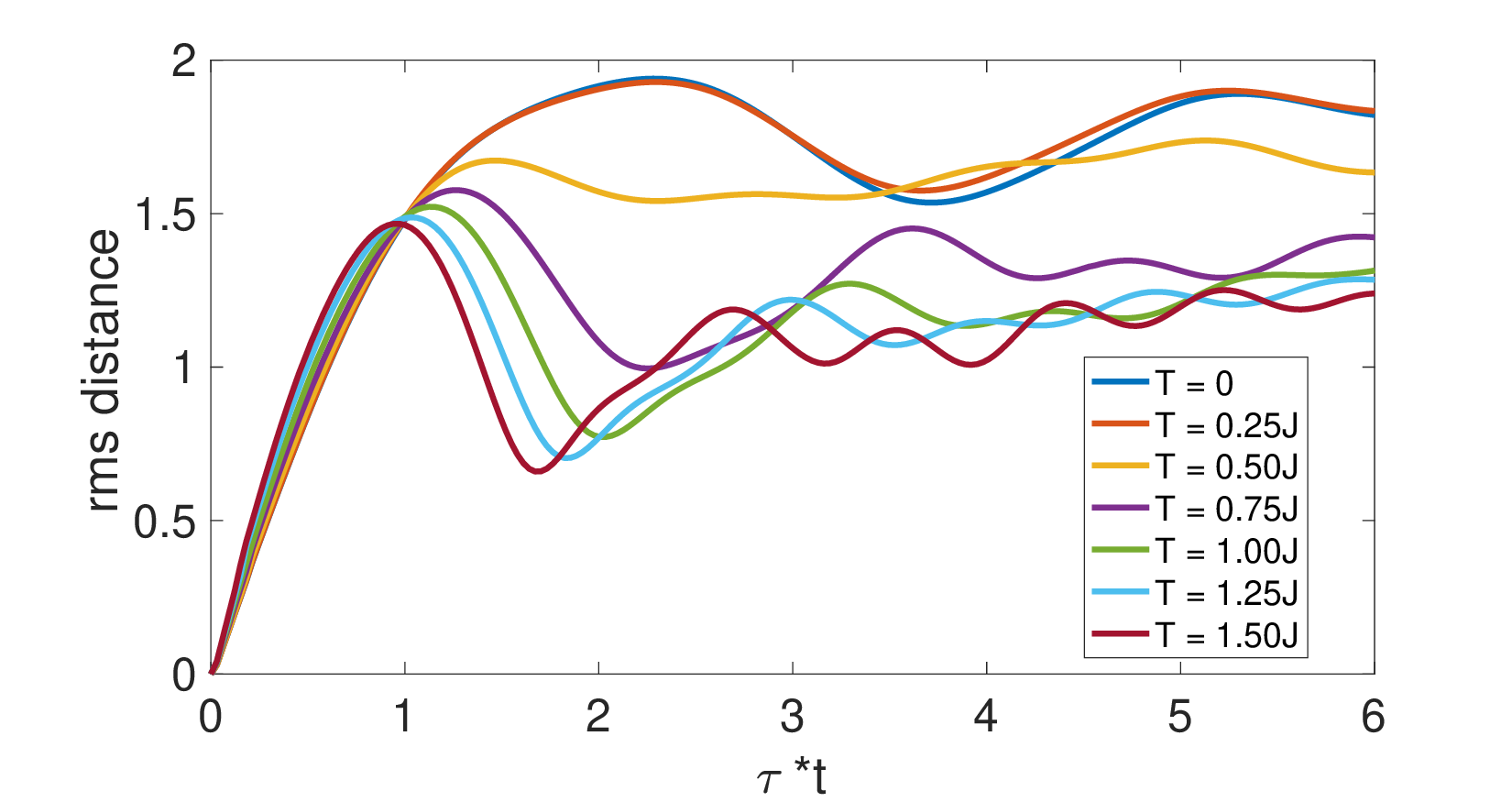}
	\caption{Evolution of the rms distance $d_{\rm rms}(\tau)$ of the hole for $J/t =0.2$ and several temperatures indicated in the legend.}
	\label{Fig2}
\end{figure}

Fig.~\ref{Fig2} illustrates how temperature modifies the $d_{\rm rms}(\tau)$ evolution for $J/t = 0.2$, a typical value of $J/t$ for high-$T_{\rm c}$ cuprate superconductors \cite{Wen}. The three-stage hole-motion scenario remains valid, though with significant temperature-induced modifications at each stage. As $T$ increases,
 $d_{\rm rms}(\tau)$ exhibits a shorter stage (i), shows earlier reversion of the hole motion and closer returning to the IHS during stage (ii) and demonstrates a slow expansion (quasi-localization) at shorter distances from the IHS during  stage (iii). These all phenomena signify the magnetic polaron effect, which is responsible for the effective reduction of the hole-hole couplings with temperature and the ensuing suppression of the hole tunneling. Semi-quantitatively, the coupling  decreases $\sim \sum_{\bold q' \bold q''}\xi^{\bold k \bold q}_{\bold q' \bold q''}\exp(-\coth\{(\omega_{\bold q'}+\omega_{\bold q''})/(2T)\})$ where $\xi^{\bold k \bold q}_{\bold q' \bold q''}$ are determined by the parameters of the model Hamiltonian~\cite{Silbey85b,Yang11}.
 Interestingly, Monte Carlo simulations of the hole transport in the $t\text{-}J_z$ model related to the anisotropic spin-spin interaction in the same (strong-coupling) regime yield qualitatively similar oscillatory evolutions of $d_{\rm rms}(\tau)$ but demonstrate the opposite trend, predicting that  the long-time value of $d_{\rm rms}(\tau)$ increases with temperature \cite{Grusdt22}. This may be caused by diminished quantum interference effects in the  $t\text{-}J_z$ model, in which Heisenberg couplings between spins are replaced with Ising couplings.

\begin{figure}[htb]
\centering
\subfigure*[]{
\includegraphics[scale=0.25,trim=150 50 80 0]{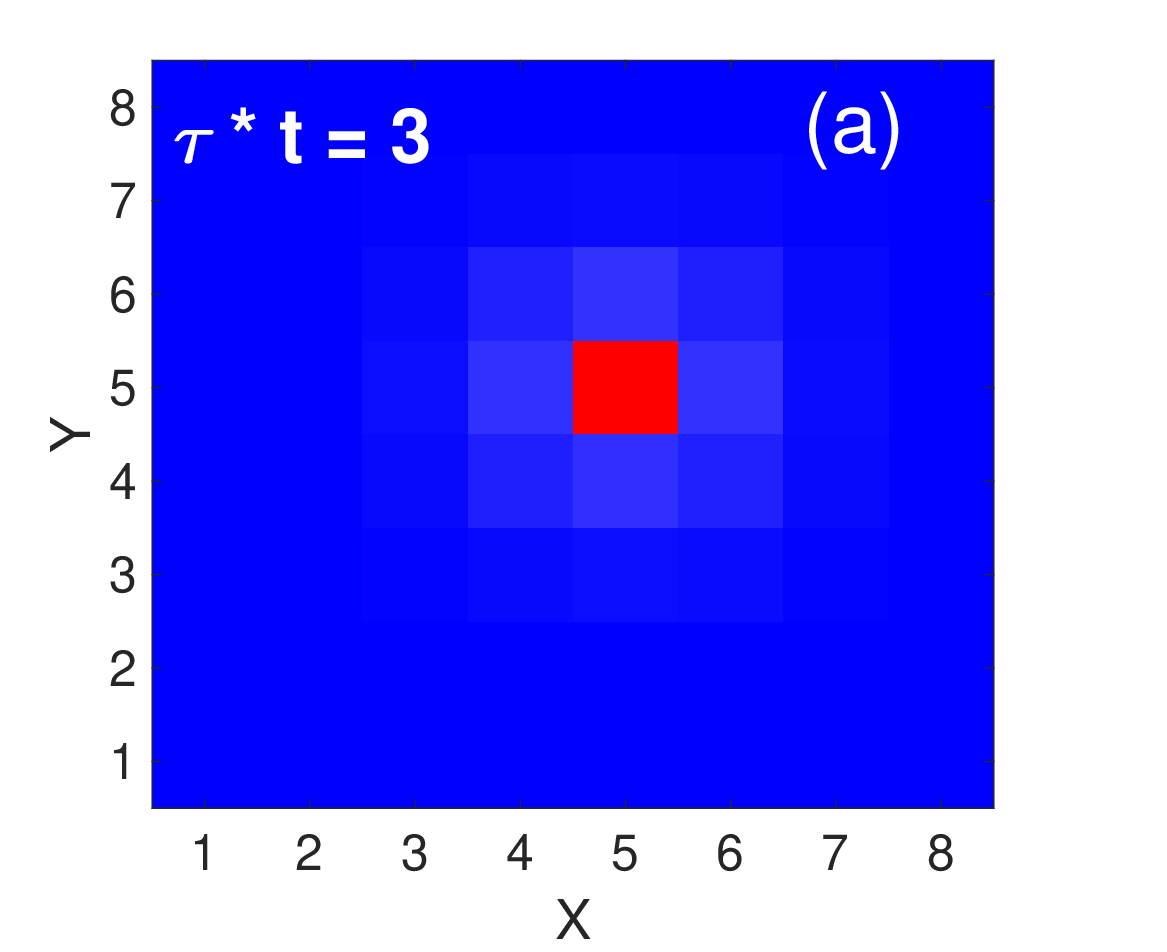}
}
\quad
\subfigure*[]{
\includegraphics[scale=0.25,trim=150 50 80 0]{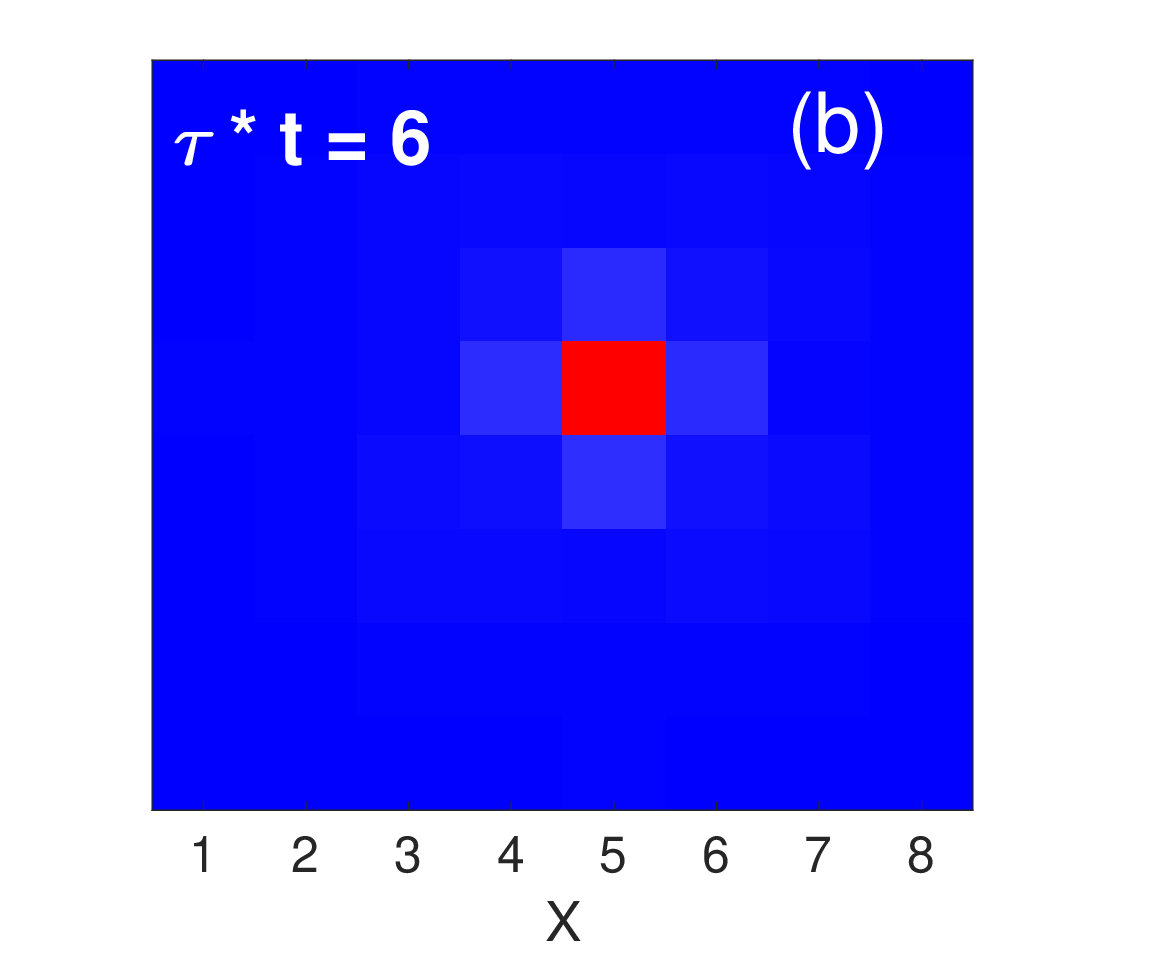}
}
\quad
\subfigure*[]{
\includegraphics[scale=0.25,trim=150 50 80 0]{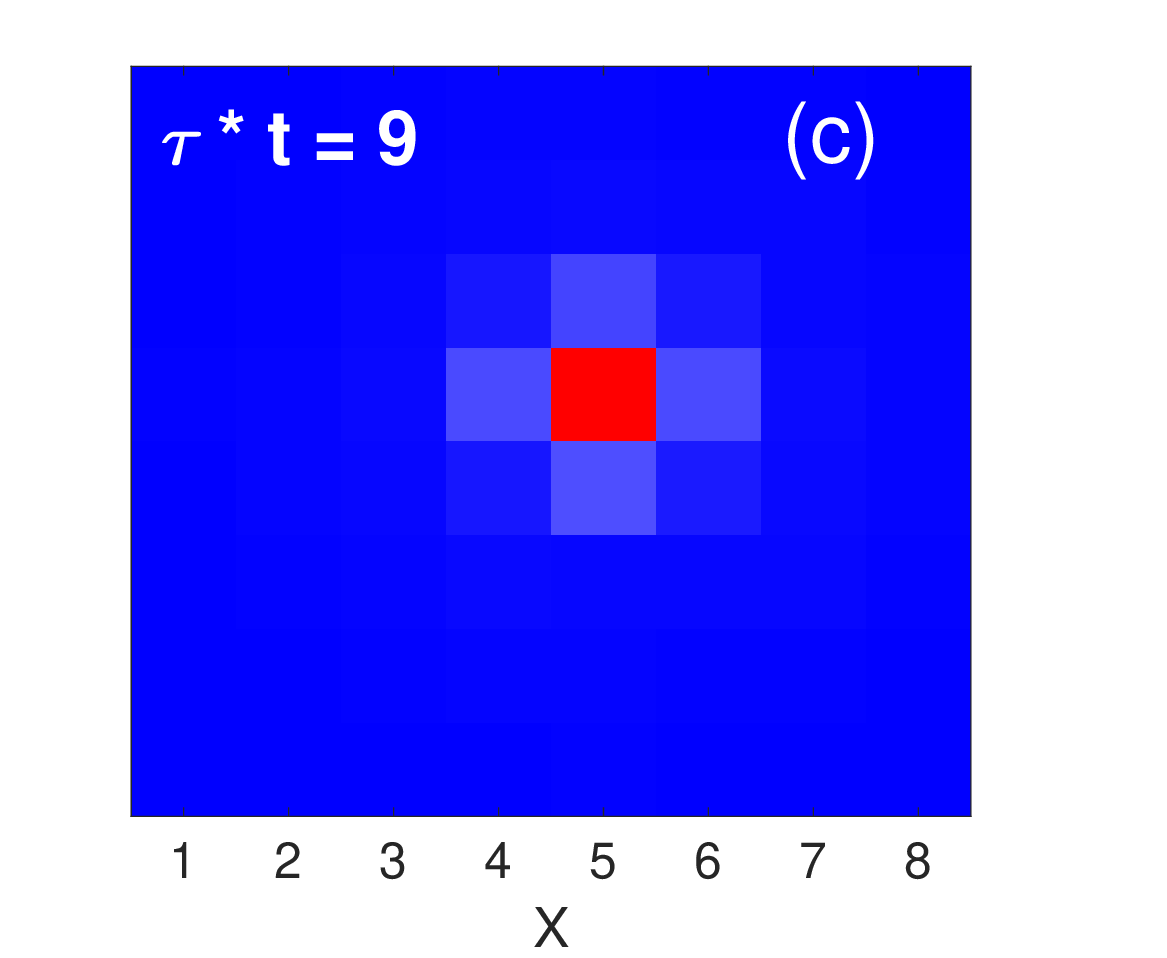}
}
\quad
\subfigure*[]{
\includegraphics[scale=0.25,trim=150 50 80 0]{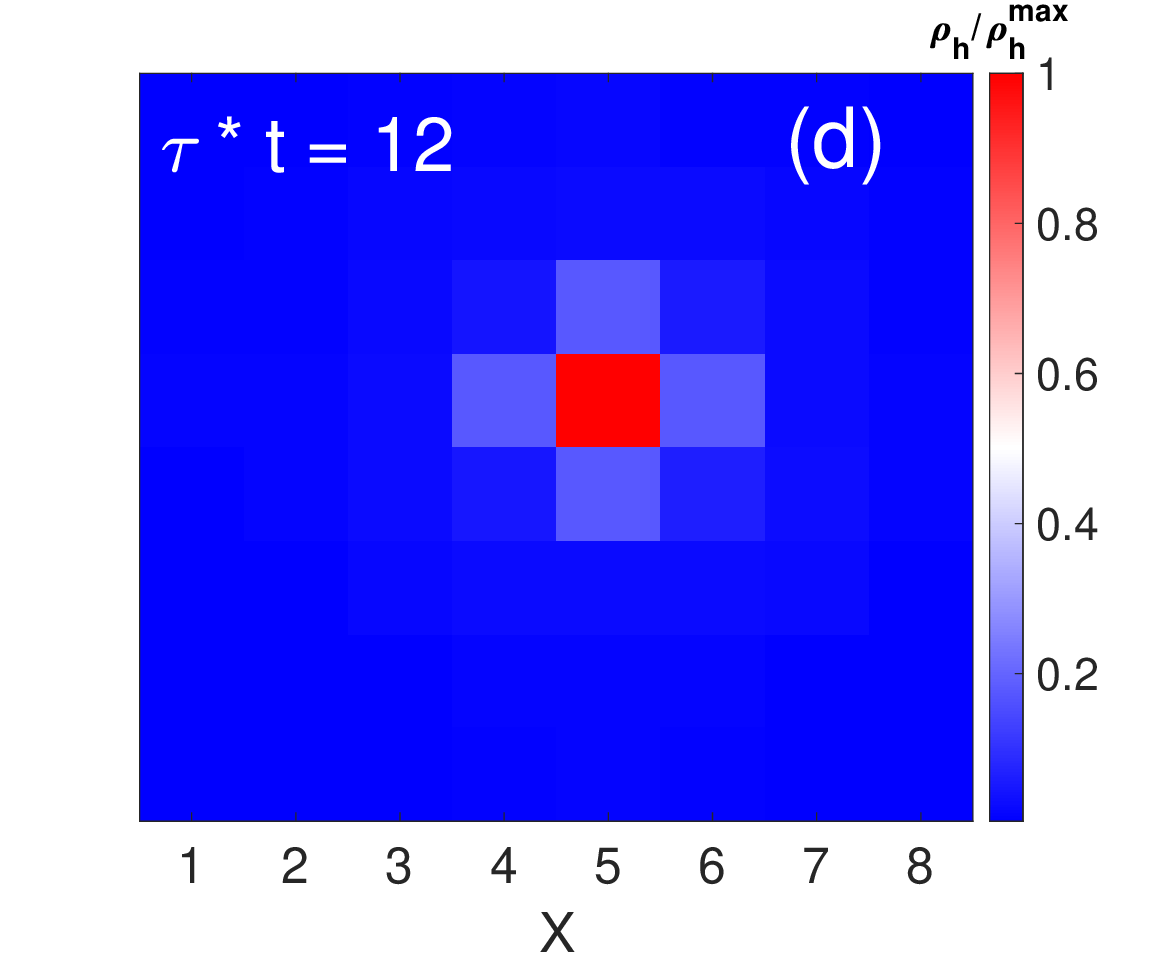}
}\\
\subfigure*[]{
\includegraphics[scale=0.25,trim=150 50 80 0]{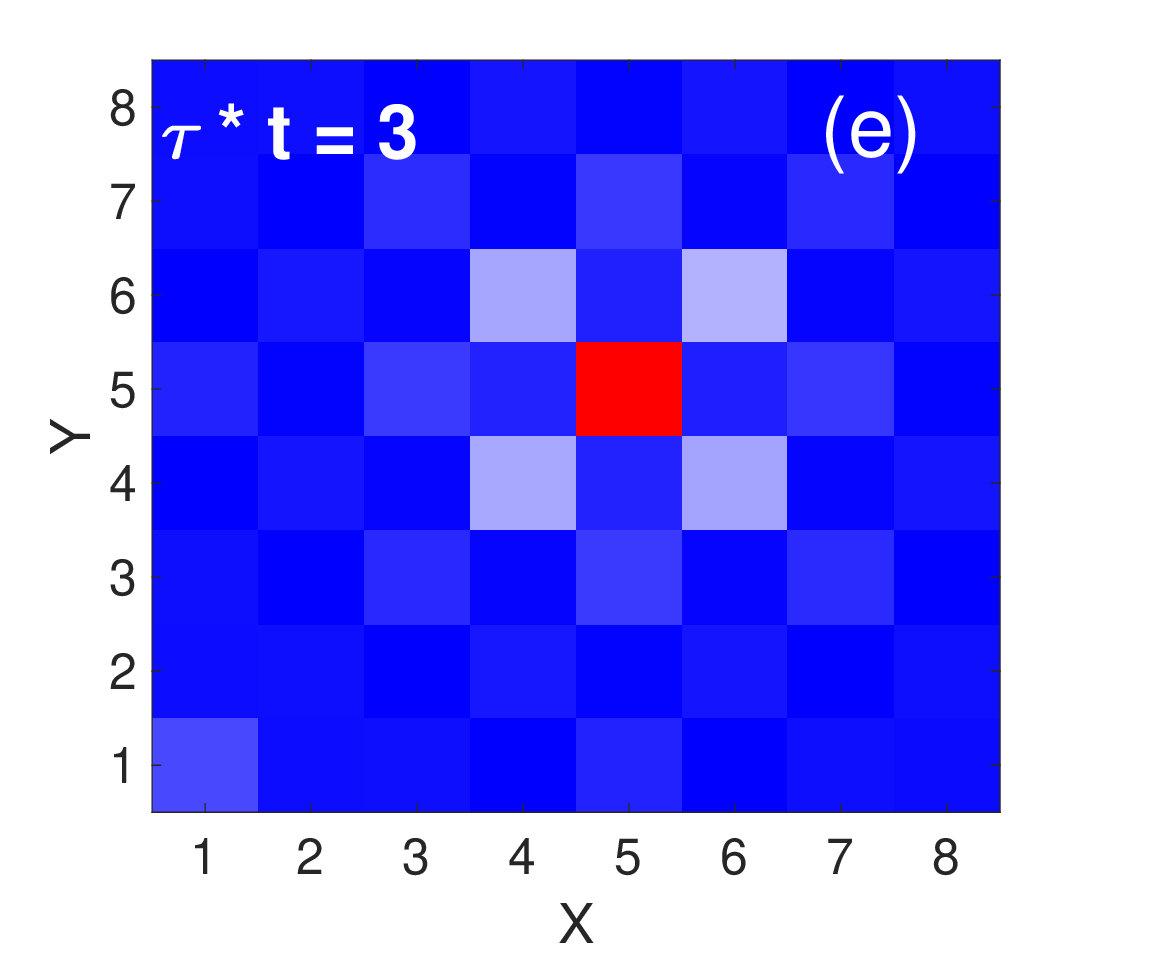}
}
\quad
\subfigure*[]{
\includegraphics[scale=0.25,trim=150 50 80 0]{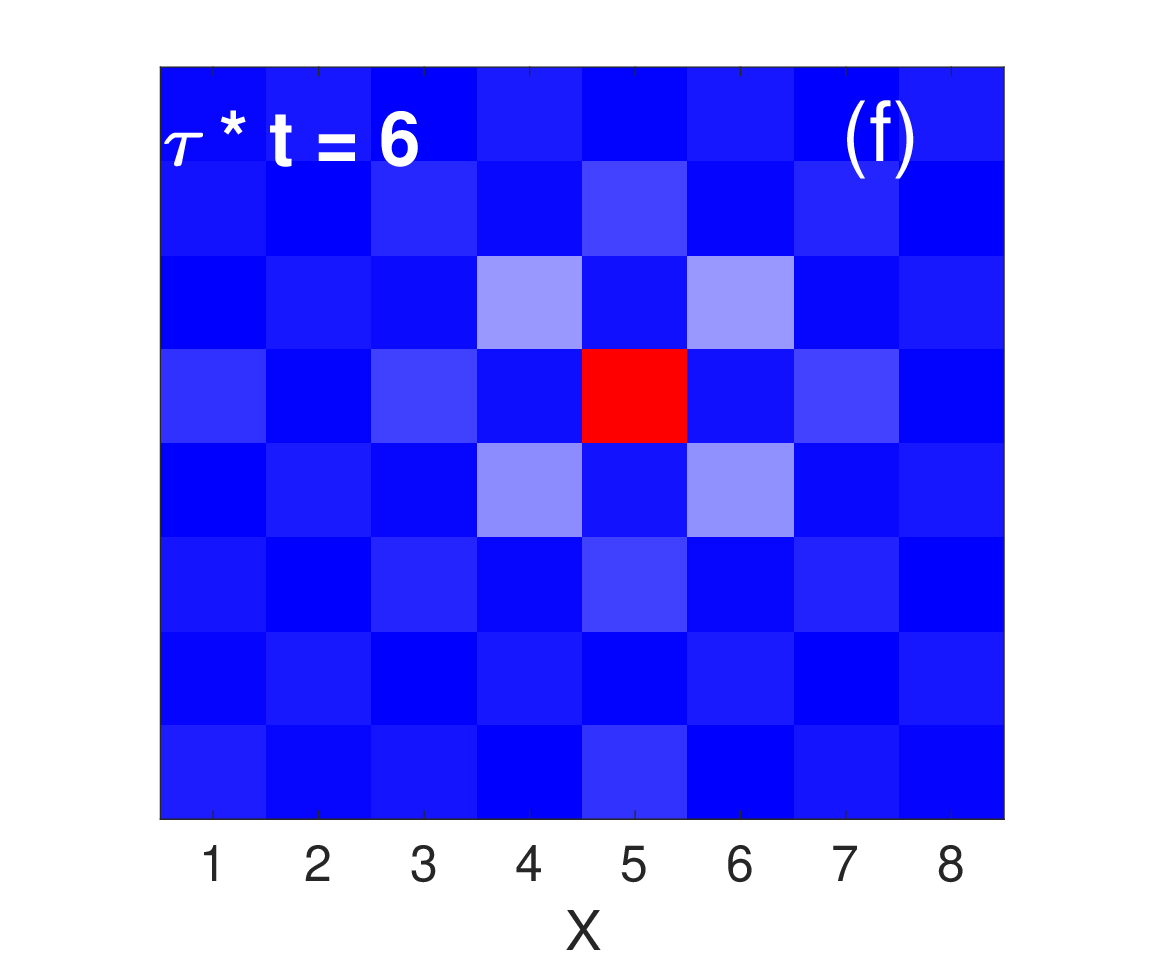}
}
\quad
\subfigure*[]{
\includegraphics[scale=0.25,trim=150 50 80 0]{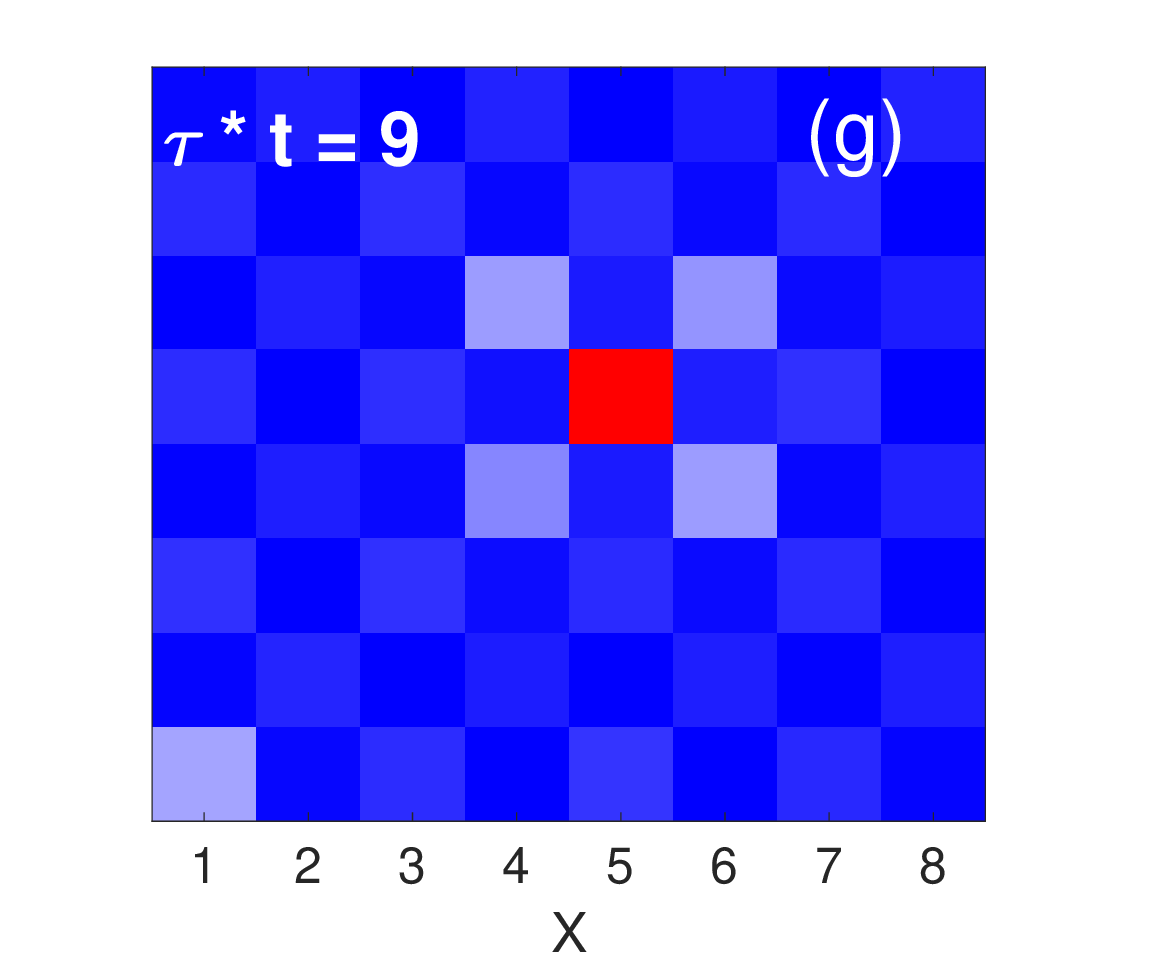}
}
\quad
\subfigure*[]{
\includegraphics[scale=0.25,trim=150 50 80 0]{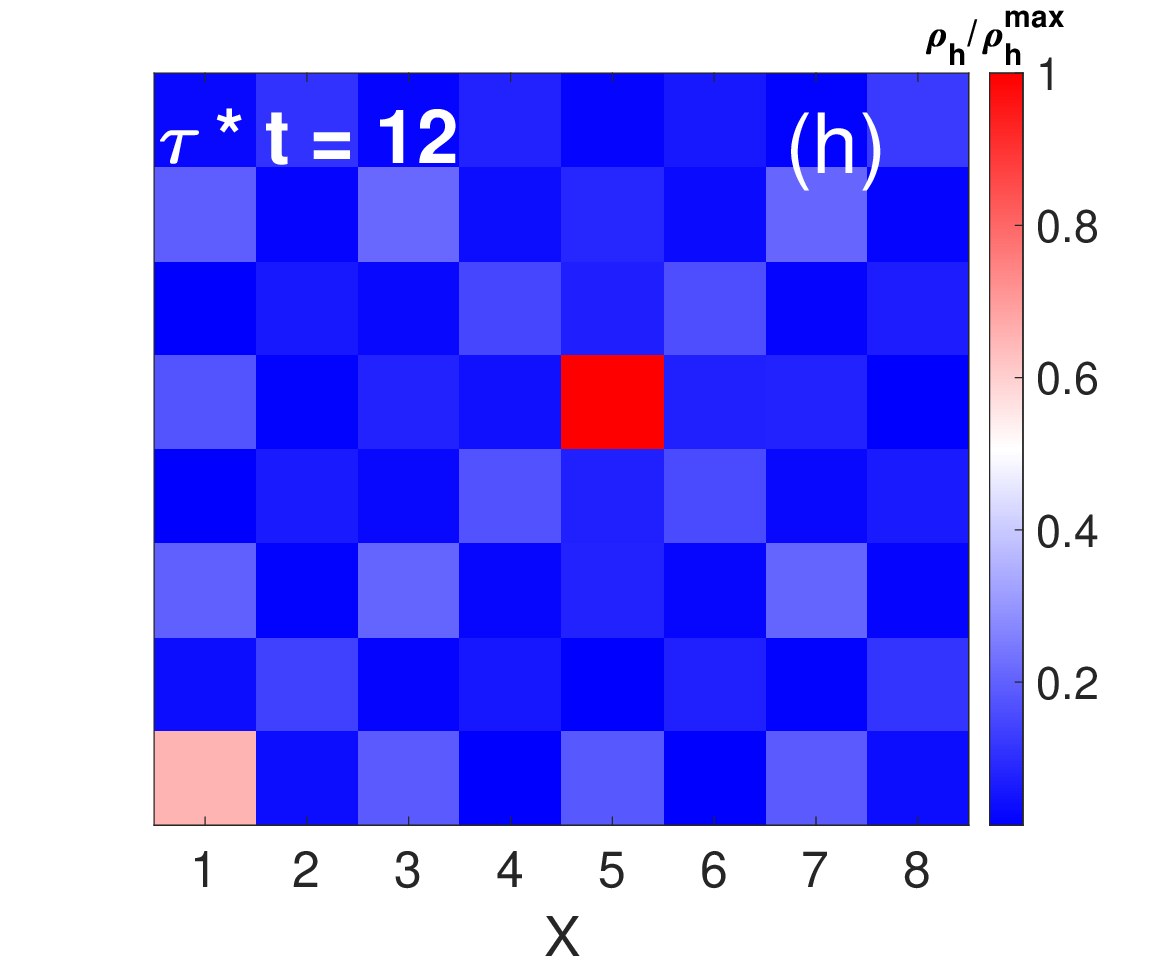}
}\\
	\includegraphics[scale=0.19,trim=160 30 30 0]{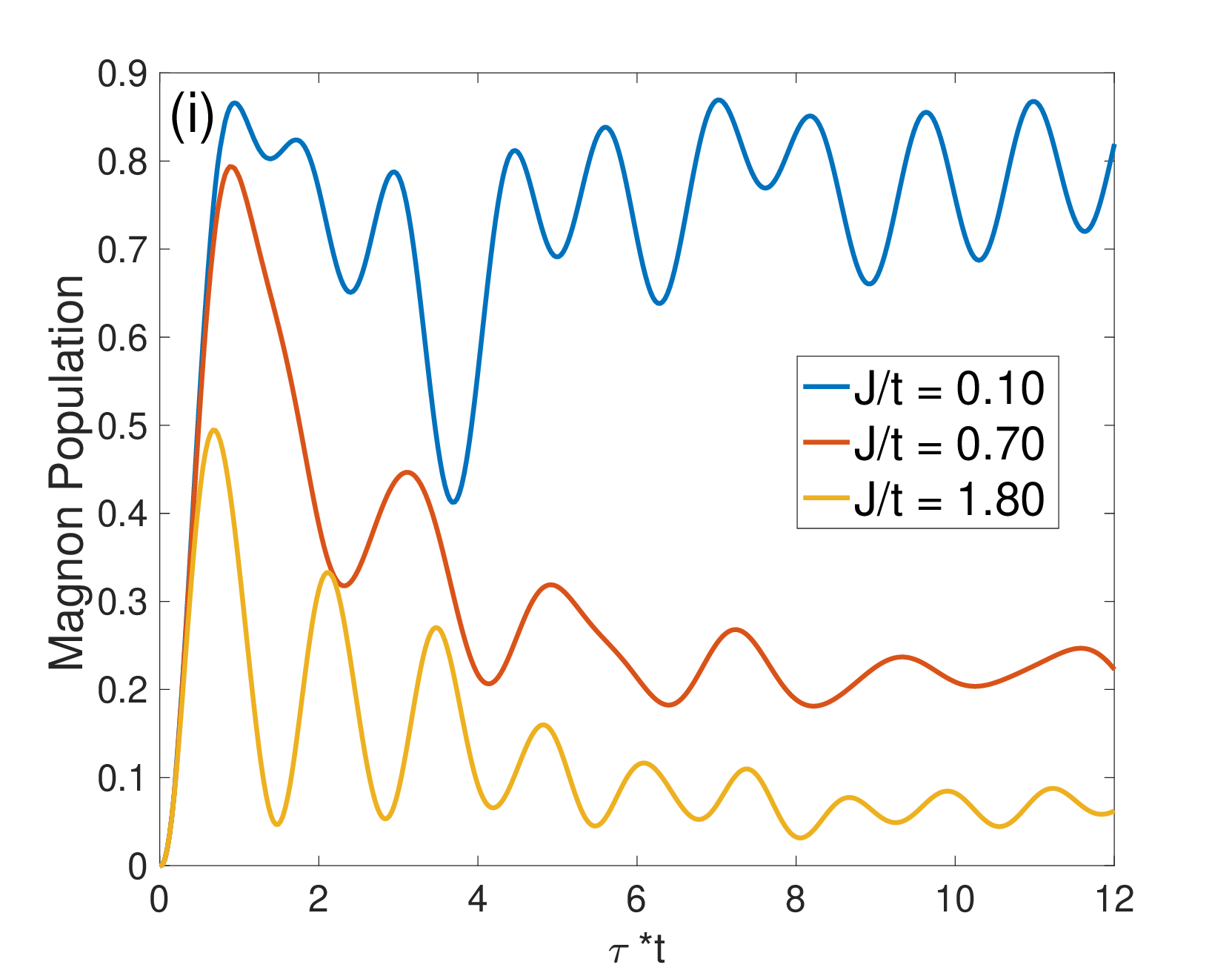}
	\includegraphics[scale=0.19,trim=50 30 0 0]{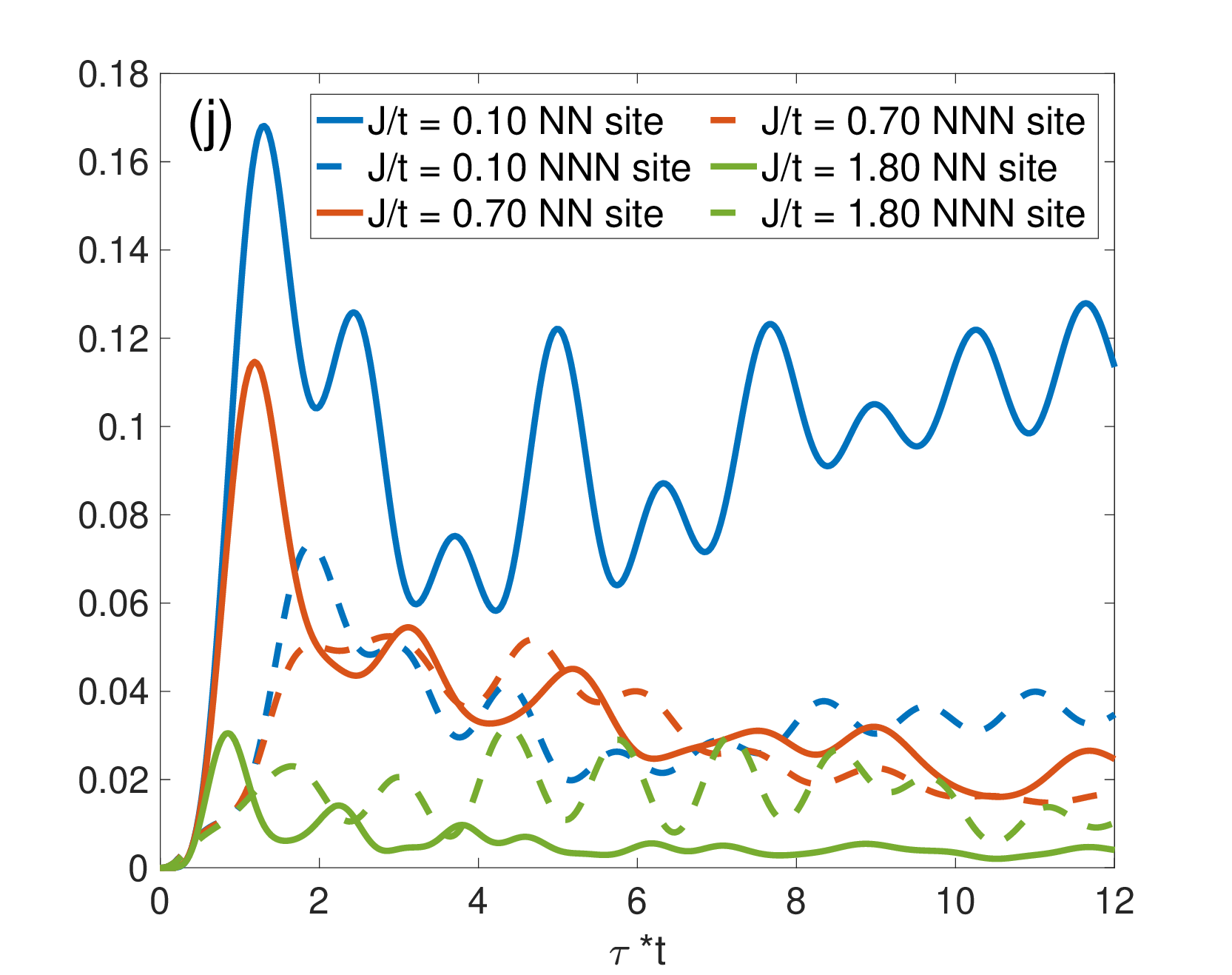}
	\includegraphics[scale=0.14,trim=100 80 200 -50]{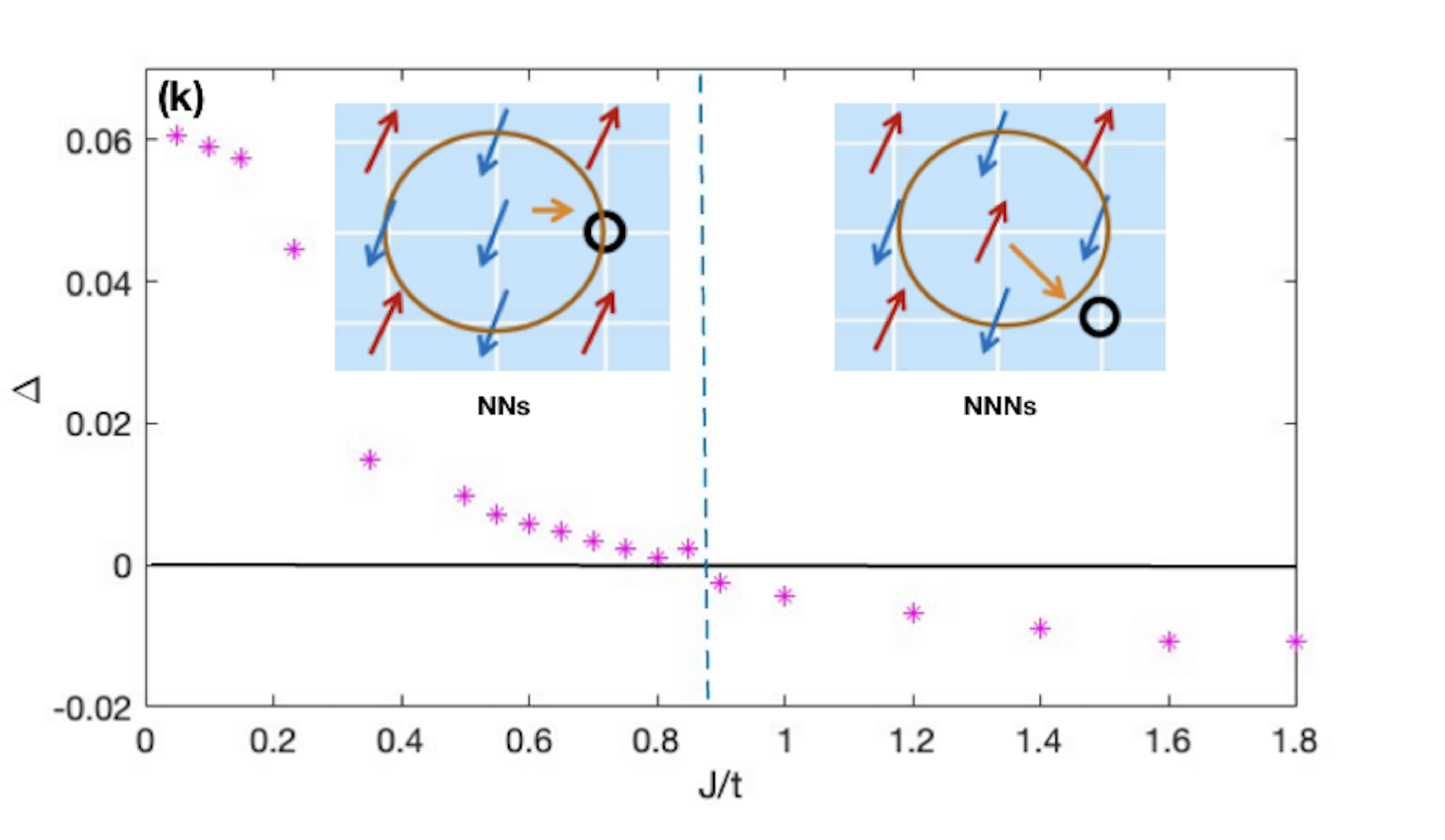}
\caption{Magnon population dynamics at zero temperature. Panels (a)-(d) and (e)-(h) give 2D views of magnon populations in the $8 \times 8$ site space for $J/t = 0.10$ and $J /t= 1.80$, respectively. Panel (i)  shows IHS magnon populations while panel (j) shows NN (full lines) and NNN (dashed lines) magnon populations for spin-spin couplings $J/t$ indicated in the legend. (k) The difference $\Delta$ between the averaged NN and NNN magnon populations (Eq.~(\ref{D})) as a function of spin-spin coupling $J/t$.}
\label{Fig3}
\end{figure}

The mDA methodology is capable of delivering hole and spin evolutions in the site space as well as in the momentum space -- all these observables can be directly computed through mDA wave functions of Eq. (\ref{D2}) by adopting basis set transformations (see Supporting Information). For example, a  direct visualization of the transport pathways is provided by  magnon populations in the site basis, which are calculated as $\rho_m(\bold d, \tau) =\langle D_2^M(\tau)|b_{\bold d}^{\dagger} b_{\bold d} |D_2^M(\tau) \rangle$, where $b_{\bold  d}^{\dagger}$ and $b_{\bold d}$ are obtained from $b_{\bold  q}^{\dagger}$ and $b_{\bold q}$ by the  unitary rotation.  $\rho_m(\bold d, \tau)$ can be used for the construction of several coarse-grained magnon populations $\rho_m(d,\tau)$, where  $d=0, \, 1, \, \sqrt{2}, \, 2$  correspond to IHS, NNs, NNNs, and SNNs, respectively.

Magnon dynamics at zero temperature is presented in Fig.~\ref{Fig3}.
Panels (a)-(d) and (e)-(h) elucidate mechanisms of magnon diffusion in real space, giving stroboscopic view of the site populations at $\tau =3j/t$, $j=1,2,3,4$. For weak spin-spin interactions (panels (a)-(d))  magnons spread only over NNs on the timescale $\tau =12/t$. The magnon confinement within the NN area is caused by the hole dressing, as discussed above. Strong spin-spin coupling (panels (e)-(h)) facilitates magnon movements, allowing for magnon diffusion through NNs at intermediate times (panels (e, f)), and to NNNs  at longer times (panels (g, h)).
Panels  (i) and (j) provide another perspective of the diffusion process, highlighting competition between the IHS,  NN  and NNN  populations.
Panel (i) shows IHS populations $\rho_m(0, \tau)$ which, after a short time $\sim \tau_{(i)}$,  attains values around 0.85, 0.8, and 0.5  for $J/t = 0.10$, $0.70$, and $1.8$, respectively. Then $\rho_m(0, \tau)$ evolves in the oscillatory manner, mirroring magnon population exchanges between the IHS, NN, and NNN areas. Owing to the hole dressing for  $J / t = 0.1$, $\rho_m(0, \tau)$  remains high and even grows slowly at  $\tau >5/t$. As demonstrated by the magnon populations in panel (j), for intermediate ($J / t = 0.7$) and strong ($J / t = 1.8$) spin-spin interactions, $\rho_m(0, \tau)$ quenches which is tantamount to the population transfer to the NN and NNN areas. For  weak spin-spin interactions ($J/t = 0.1$), NN magnon populations remain larger than their NNN counterparts up to $\tau  =12/t$, signifying magnon confinement within the NN domain. For $J / t = 0.7$, magnon populations in the NNN area surpass those in the NN area up to  $\tau \approx 2/t$, while the two populations become comparable at longer times. For strong spin-spin interactions ($J /t = 1.8$), on the other hand, the NNN populations outgrow their NN counterparts after  $\tau \approx 1/t$.
The physical picture of the magnon transport outlined above is further corroborated by panel (h), which displays the averaged difference between the NN and the NNN populations,
\begin{equation}\label{D}
\Delta=\tau_f^{-1}\int_0^{\tau_f} d \tau \left(\rho_m(1, \tau)-\rho_m(\sqrt{2}, \tau)\right) ,
\end{equation}
as a function of the spin-spin coupling $J/t$, where $\tau_f=12/t$. The NN populations dominate for small $J/t$, with $J/t \approx 0.88$ marking a crossover, above which the NNN populations take over.

\begin{figure}[htb]
\centering

\subfigure*[]{
	\includegraphics[scale=0.23,trim=150 30 100 0]{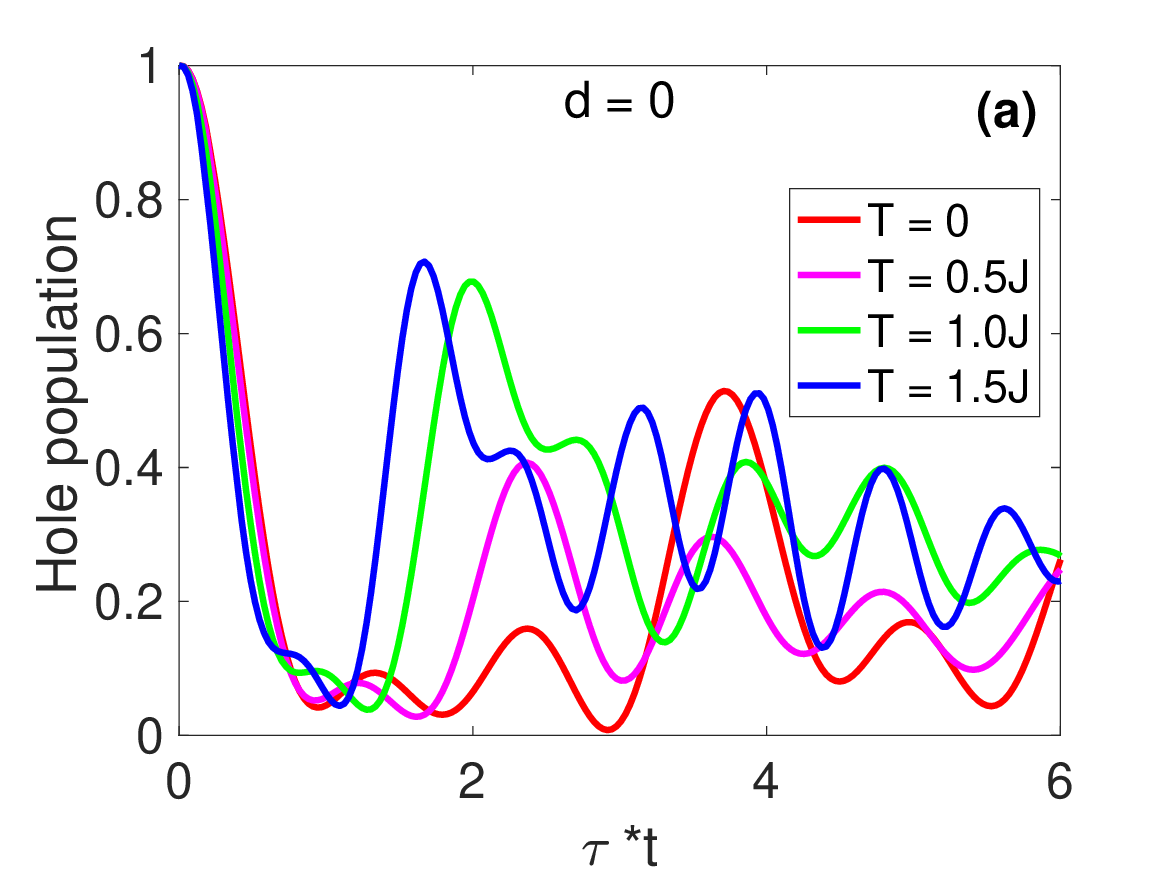}
}
\quad
\subfigure*[]{
	\includegraphics[scale=0.23,trim=90 30 100 0]{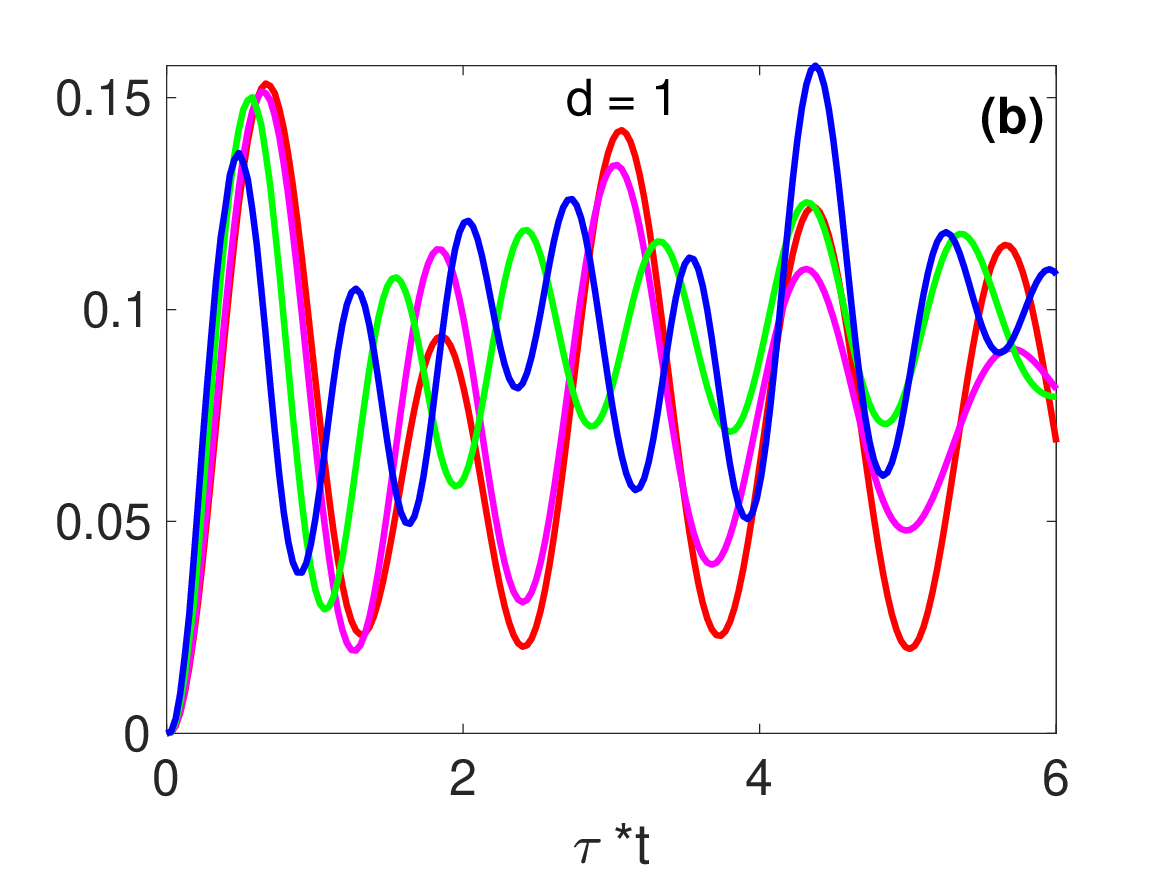}
}
\quad
\subfigure*[]{
	\includegraphics[scale=0.23,trim=90 30 100 0]{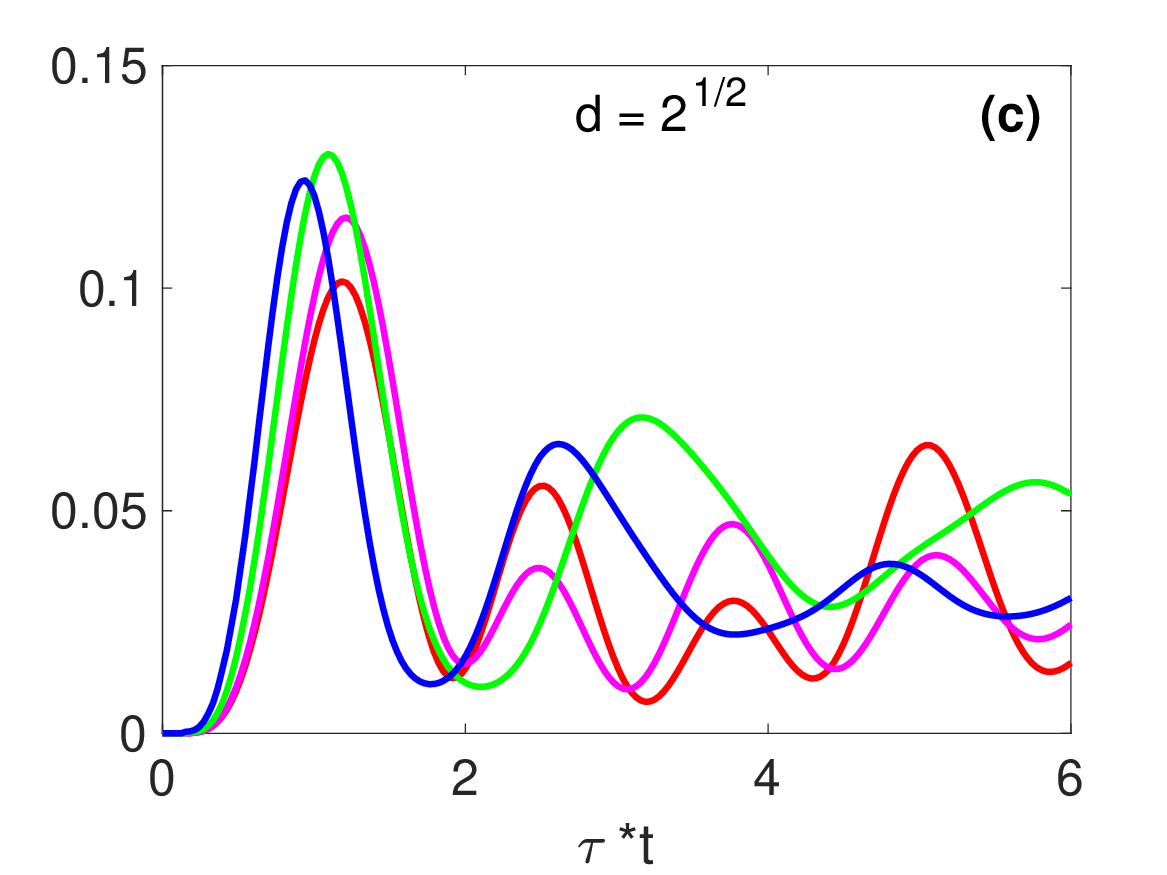}
}
\quad
\subfigure*[]{
	\includegraphics[scale=0.23,trim=90 30 100 0]{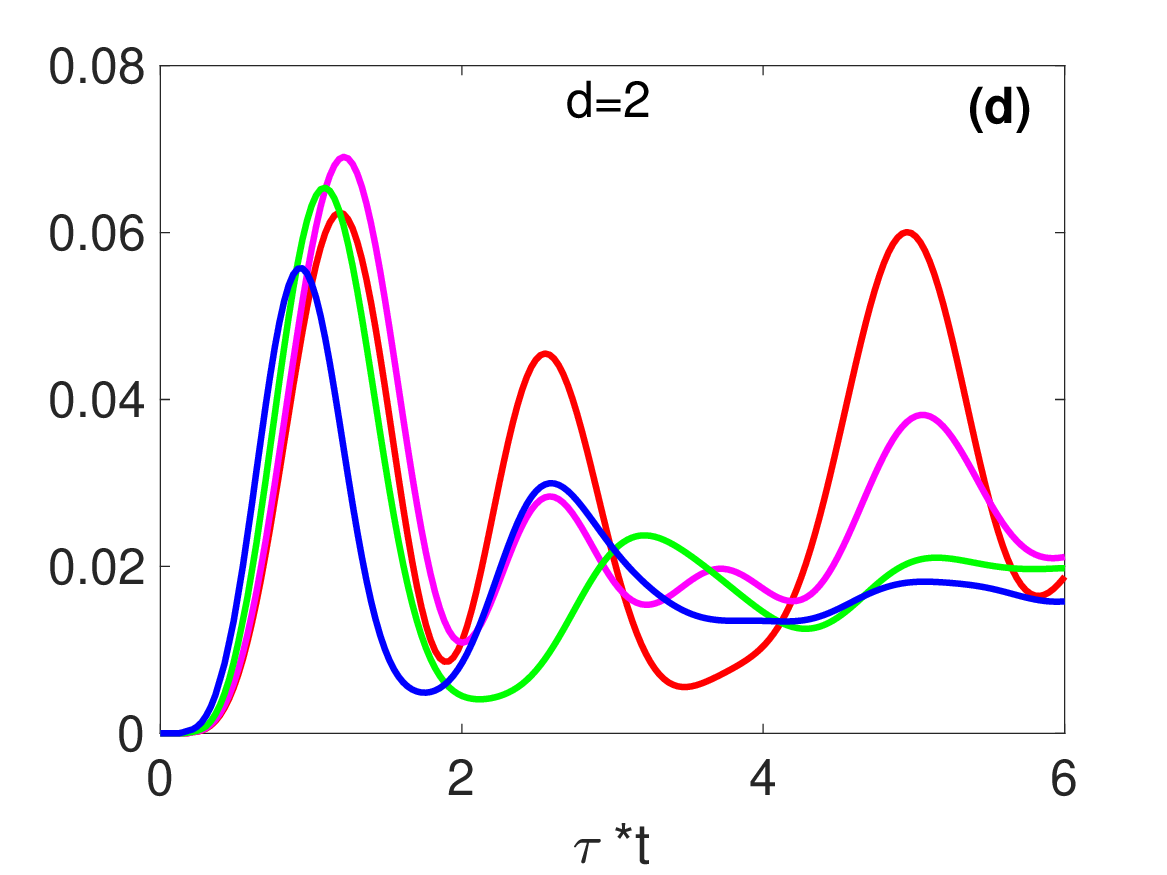}
}\\
\subfigure*[]{
	\includegraphics[scale=0.23,trim=140 30 70 53]{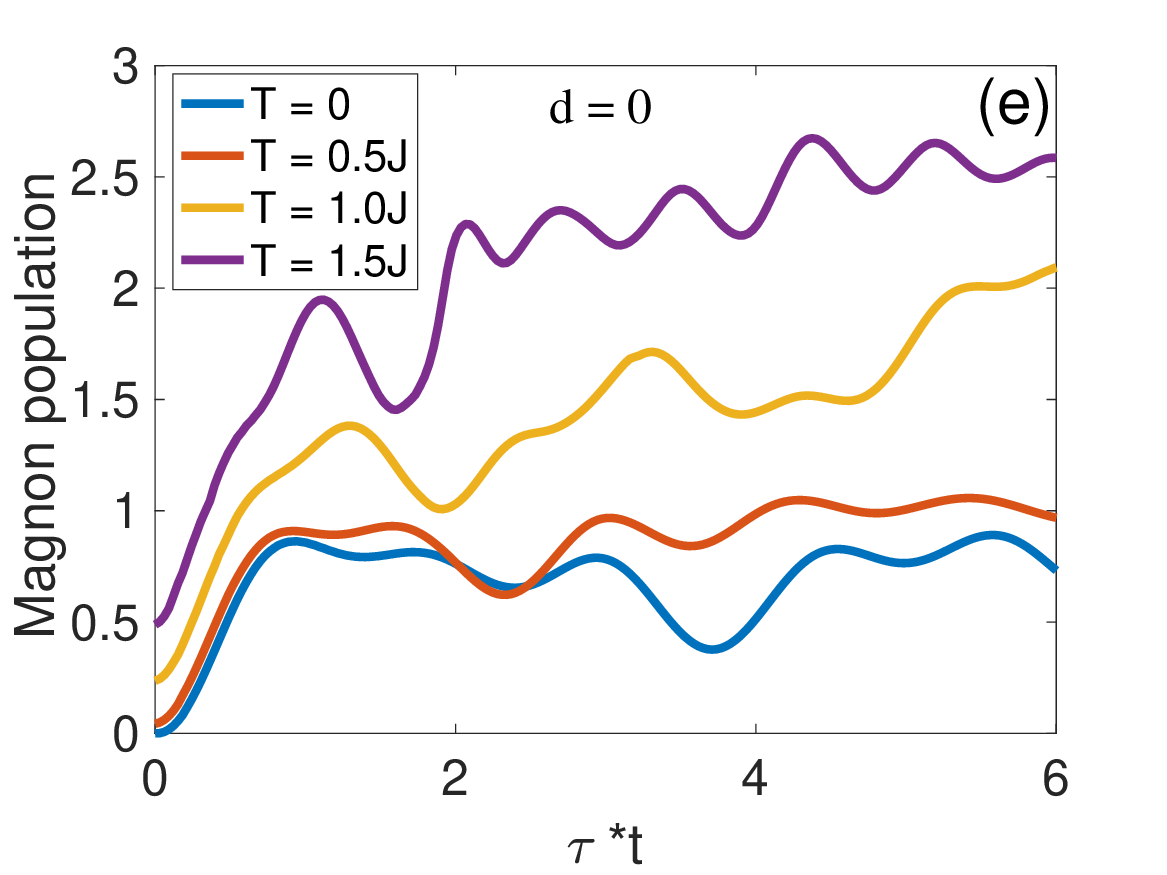}
}
\subfigure*[]{
	\includegraphics[scale=0.23,trim=70 30 100 53]{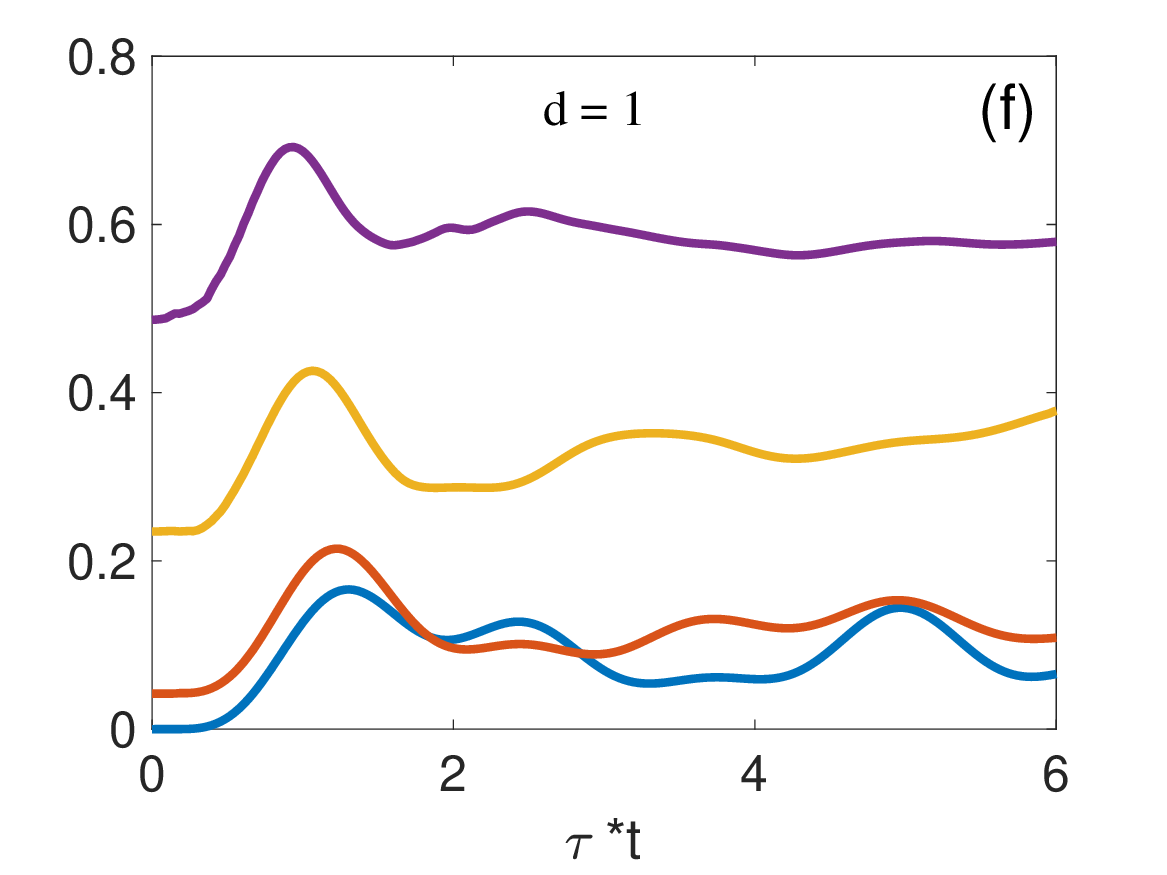}
}
\quad
\subfigure*[]{
	\includegraphics[scale=0.23,trim=90 30 100 53]{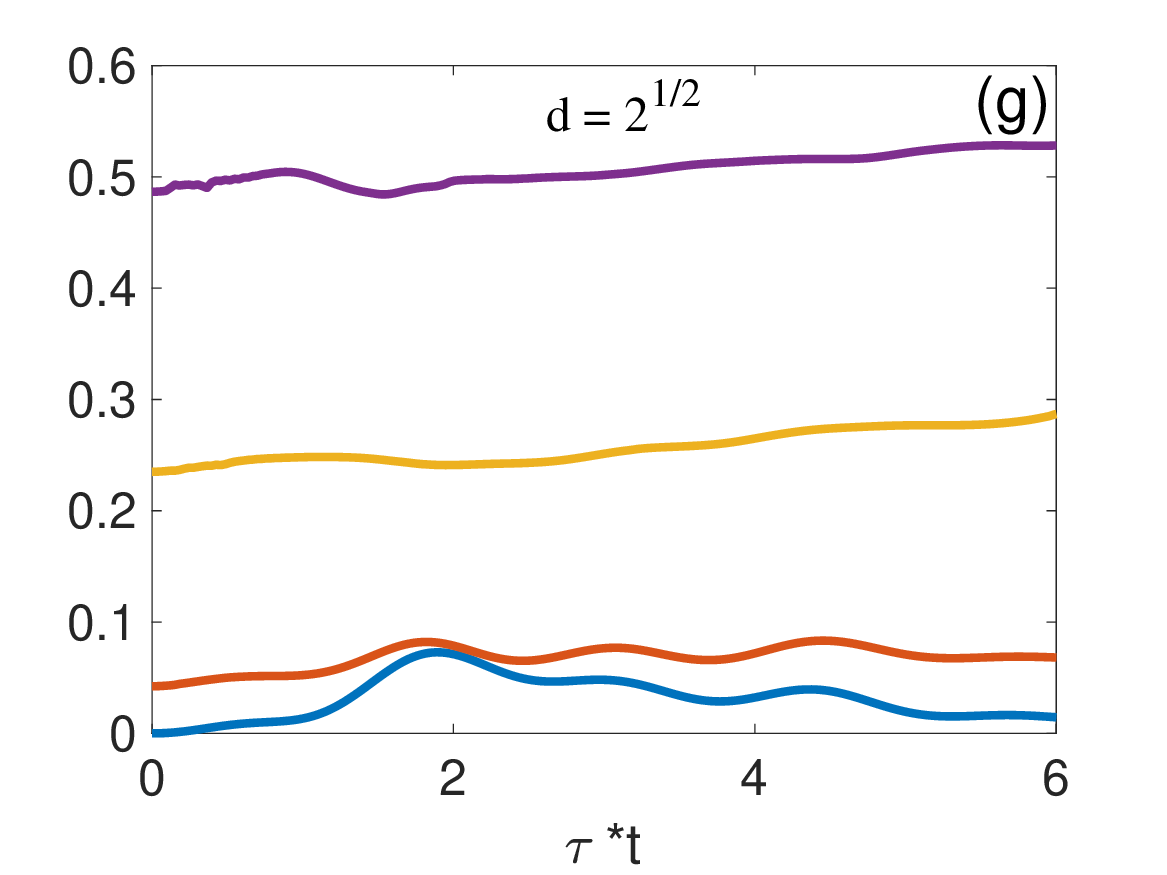}
}
\quad
\subfigure*[]{
	\includegraphics[scale=0.23,trim=90 30 100 53]{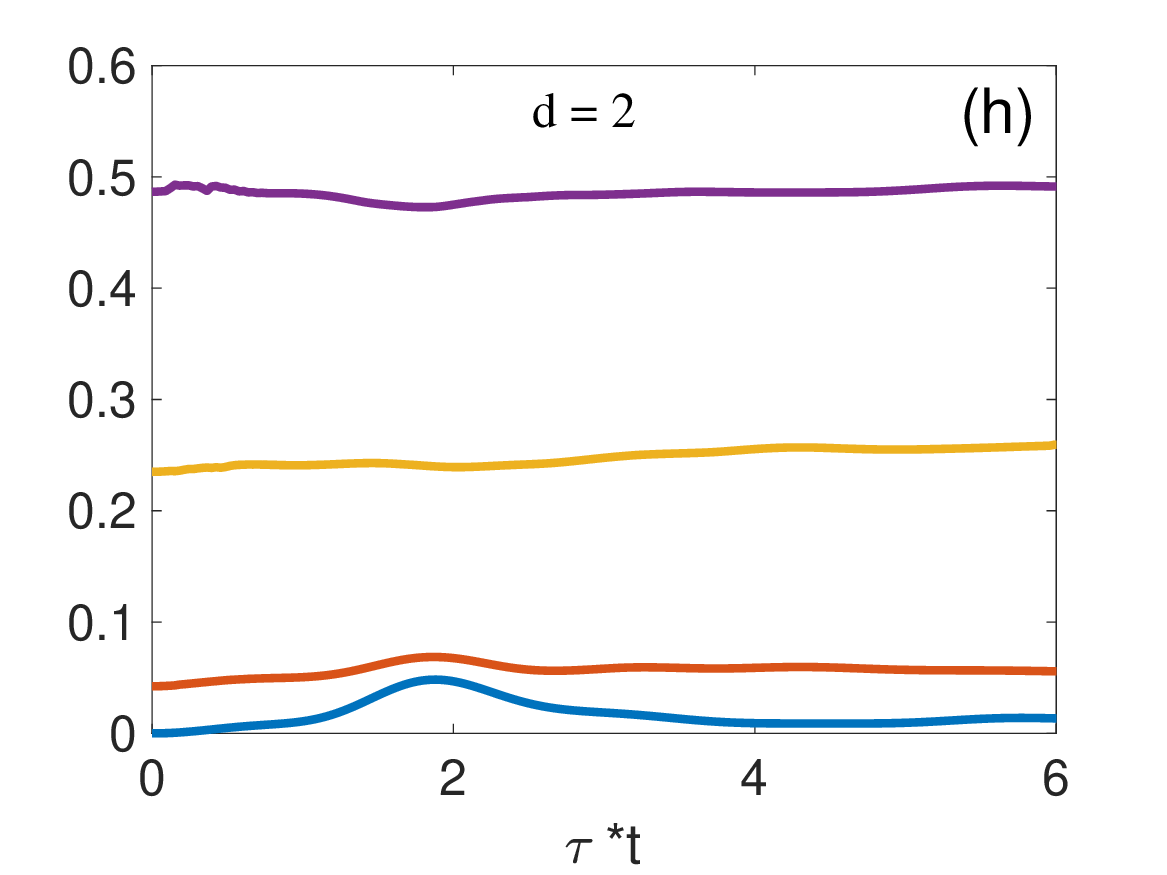}
}
\\

\caption{Hole (panels (a-d)) and magnon (panels (e-h)) populations in the IHS, NNs, NNNs, and SNNs areas for elevated temperatures and $J/t = 0.2$.  }
\label{Fig4}
\end{figure}

\begin{figure}[htbp]
	\centering
	\includegraphics[scale=0.45]{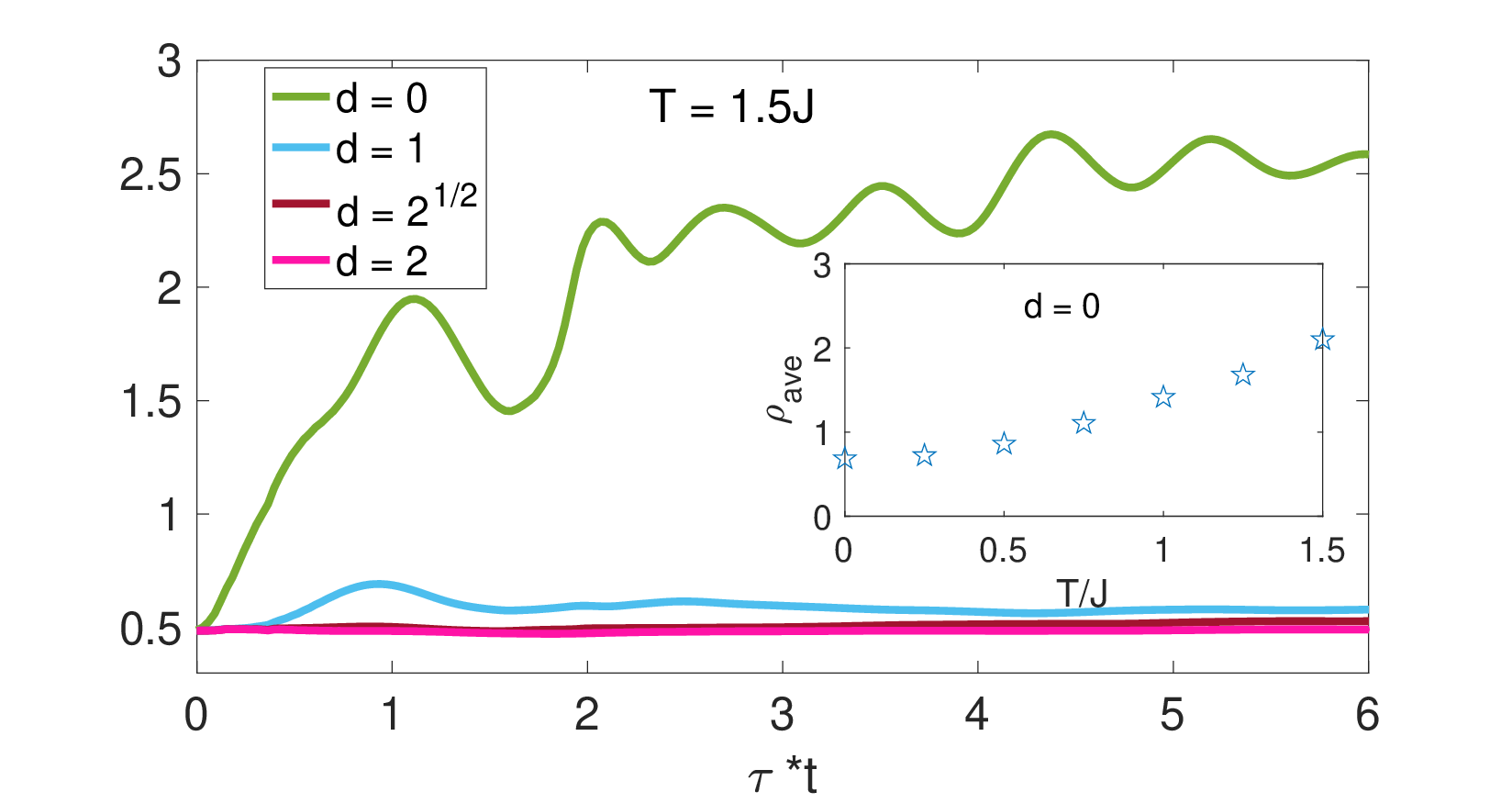}
	\caption{ Magnon populations  in the IHS, NNs, NNNs, and SNNs areas for $T/J = 1.5 $ and $J/t = 0.2$. Inset shows the average IHS magnon population vs. temperature. }
	\label{Fig5}
\end{figure}

The evolution of hole populations $\rho_h(d, \tau)$ (up row) and magnon populations $\rho_m(d, \tau)$ (bottom row) in time is elucidated in Fig.~\ref{Fig4} for elevated temperatures. From left to right, the panels correspond to IHS ($d=0$), NN ($d=1$), NNN ($d=\sqrt{2}$) and SNN ($d=2$).
We begin with general observations. Hole populations exhibit pronounced oscillations at $T=0$ (which correlates, e.g., with the tensor-network calculations \cite{Cirac20}) and elevated temperatures. Remarkably, the amplitudes and periods of oscillations in the IHS (a) and NN (b) hole populations increase with temperature. This is a manifestation of the $T$-enhanced hole-hole coupling caused by the increasing number of thermally-activated magnon states controlling the coupling. In the effective Rabi-model picture, the enhanced coupling  amplifies Rabi frequency governing hole-hole population transfer. Similar (though significantly mitigated) effects are observed for magnon populations (e, f). For NNN and SNN hole (c, d) and magnon (g, h) populations, there is no clear trend in the  $T$-induced modifications  of the amplitudes and the oscillation periods, as long-range, magnon-assisted transfer of holes is inevitably accompanied by multiple forward-backward scatterings in the IHS, NN, NNN, and SNN areas which induces dephasing.

The IHS hole populations in Fig.~\ref{Fig4}(a) exhibit pronounced recurrences, the amplitude (the period) of which increases (decreases) with temperature -- in full agreement with the behavior of rms distances $d_{\rm rms}(\tau)$ in Fig.~\ref{Fig2}.
The NN hole population dynamics (panel (b)) shows a similar pattern. NNN (c) and SNN (d) hole populations exhibit irreversible $T$-induced quenching.
Magnon populations (e)-(h), on the other hand, reveal two important general trends.
First, $\rho_m(d, \tau)$  increases with temperature, mirroring thermal activation of magnons.
Second, except for $\rho_m(0, \tau)$, magnon populations rapidly attain their steady-state values, notably at higher $T$. This is  attributed to destructive quantum interference among multiple thermally-populated magnon states.

The typical values of magnon populations $\rho_m(d, \tau)$, in the IHS ($d=0$), NN ($d=1$), NNN ($d=\sqrt{2}$) and SNN ($d=2$) areas at $T=1.5J$, are depicted in Fig.~\ref{Fig5}. Clearly,  the IHS population is higher than the remaining three populations taken together. The inset shows that the averaged IHS magnon population $\rho_{ave}= \tau_f^{-1}\int_0^{\tau_f} d \tau \rho_m(0, \tau)$  increases monotonically with $T$. This can be interpreted as thermally-induced frustration of the spin alignment and correlates with the decrement of  saturation magnetization at elevated temperatures \cite{Prya, Olss}.

In summary, we have elaborated a computationally-efficient, numerically-accurate mDA-TFD methodology for simulating hole-magnon dynamics of the $t\text{-}J$ model at finite temperatures. This permits one to scrutinize previous SCBA predictions~\cite{knak} while offering, for the first time, finite-temperature calculations of detailed magnon dynamics as a response and a facilitator to the hole motion. Our study also uncovers pronounced $T$-dependence of the magnon and hole populations, pointing to the feasibility of potential thermal manipulation and control of hole dynamics. Time evolutions of magnon populations computed here can be measured by the micro-focus Brillouin light scattering (BLS) spectroscopy~\cite{Divi,Demi}. Furthermore, our methodology can be applied not only to the calculation of steady-state angular-resolved photoemission spectra~\cite{Bohr,Bohr20},
but also to the simulation of femtosecond terahertz pump-probe and other nonlinear signals that have been used for the characterization of AFM materials  \cite{Wang23}, as it has been repeatedly shown that the mDA method is a reliable approach to nonlinear time- and frequency-resolved spectra of various material systems \cite{zhao1,zhao2,sun3,sun4}.
It seems especially promising to compare predictions of our simulations on magnon dynamics with measurements of femtosecond terahertz 2D spectroscopy, which is capable of monitoring inter-magnon temporal and spacial correlations \cite{Nelson17,Taka23}.
In addition, the mDA-TFD framework  holds the potential to extend its application to an analysis of magnon polaritons \cite{Kato23}, AFM  bilayers \cite{knak23a}, and  nonequilibrium dynamics of multiple holes~\cite{knak23, koeps, chiu} in strongly interacting lattice models. Such extensions could foster a more nuanced comprehension of the interplay between holes, magnons, and polarons, and further illuminate fascinating phenomena such as {\emph{d}}-wave Cooper pairs, stripe phases, and {\emph{d}}-wave superconductivity.

\section*{Data Availability Statement}
The data that support the findings of this study are available from the corresponding author upon reasonable request.

\section*{Supporting Information}
The following file is available free of charge.
\begin{itemize}
  \item supp\_v5: The thermo-field dynamics approach, variational equations of motion based on the multi-D2 Ansatz for the $t\text{-}J$ model
\end{itemize}

\section*{ACKNOWLEDGMENTS}
The authors gratefully acknowledge the support of the Singapore Ministry of Education Academic Research Fund (Grant No. RG87/20). K.~Sun would also like to thank the Natural Science Foundation of Zhejiang Province (Grant No.~LY18A040005) for partial support. M. F. G. acknowledges the support of Hangzhou Dianzi University through startup funding.


\begin{thebibliography}{999}
\bibitem{footnote} TFD-TT machinery is akin to the time-dependent
density matrix renormalization group \cite{White05} and other tensor network techniques \cite{Qin22,Sinha22}.
\end{thebibliography}


\clearpage 
\appendix
\section*{}
\setcounter{equation}{0}
\newcommand*\mycommand[1]{\texttt{\emph{#1}}}


\title{\textbf {\Large Supporting Information: 
Finite-Temperature Hole-Magnon Dynamics in an Antiferromagnet}
}
\renewcommand{\thepage}{S-\arabic{page}}

%
%

\section{\Large S1. The thermo-field dynamics approach}
\renewcommand{\theequation}{S\arabic{equation}}

The finite temperature effects can be simulated by introducing the additional "tilde" magnon degrees of freedom {$\tilde{\bf q}$~\cite{Y,M1,M2}}. Then the total
Hamiltonian acting in the extended $\{ \bf q\}= \bf q \oplus \tilde{\bf q}$ Hilbert space assumes the form~\cite{Borr}
\begin{eqnarray}
\overline H =  H - \sum_{ \bf q} \omega_{\bf q} \tilde b_{\bf q}^{\dagger} \tilde b_{\bf q}
\end{eqnarray}
where $\tilde b_{\bf q}^{\dagger}$ and $\tilde b_{\bf q}$ are the tilde creation and annihilation operators.
Having performed thermal Bogoliubov transformation specified by the operator
\begin{align}
	G = G^{\dagger}=-i \sum_{\bf q} \theta_{\bf q} (b_{\bf q} \tilde b_{\bf q}-b_{\bf q}^{\dagger} \tilde b_{\bf q}^{\dagger})
\end{align}
and following the prescriptions of {Refs~\cite{Borr,chen}}, we obtain the final thermo-field dynamics $t\text{-}J$ Hamiltonian
\begin{align}
 H_{\theta} &= e^{iG} \overline H e^{-iG} \nonumber\\
&= \sum_{\mathbf{q}} \omega_{\mathbf{q}} (b_{\mathbf{q}}^{\dagger} b_{\mathbf{q}} -\tilde b_{\mathbf{q}}^{\dagger} \tilde b_{\mathbf{q}})+ \frac {tz}{\sqrt{N}} \sum_{\mathbf{kq}} h_{\mathbf{k-q}}^{\dagger} h_{\mathbf{k}} \cosh(\theta_{\mathbf{q}})[(u_{\mathbf{q}} \gamma_{\mathbf{k-q}}+v_{\mathbf{q}} \gamma_{\mathbf{k}}) b_{\mathbf{q}}^{\dagger}+(u_{\mathbf{q}} \gamma_{\mathbf{k}}+v_{\mathbf{q}} \gamma_{\mathbf{k-q}})b_{\mathbf{-q}}] \nonumber \\
&+ \frac {tz}{\sqrt{N}} \sum_{\mathbf{kq}} h_{\mathbf{k-q}}^{\dagger} h_{\mathbf{k}} \sinh(\theta_{\mathbf{q}})[(u_{\mathbf{q}} \gamma_{\mathbf{k-q}}+v_{\mathbf{q}} \gamma_{\mathbf{k}}) \tilde b_{\mathbf{q}}+(u_{\mathbf{q}} \gamma_{\mathbf{k}}+v_{\mathbf{q}} \gamma_{\mathbf{k-q}}) \tilde b_{\mathbf{-q}}^{\dagger}].
\end{align}
The influence of temperature is imprinted into  $\overline H_{\theta}$ through the temperature-dependent  mixing angles
\begin{align}
	\theta_{\bf q} = {\rm arctanh}(e^{-\beta \omega_{\bf q}/2})\end{align}
which renormalize hole-magnon coupling coefficients.

\section{\Large S2. Multi-$\rm D_2$ ansatz and its variational equations}
\renewcommand{\theequation}{S\arabic{equation}}
The multi-$\rm D_2$ Ansatz~\cite{zheng,sun1,sun2,sun3,zhao1,zhao2} of  multiplicity $M$ in the thermo-field dynamics method can be constructed as
\begin{equation}\label{D2}
| D_2^{\rm M} (\tau) \rangle =  \sum_{\scriptstyle1\le n\le N \atop\scriptstyle1\le m\le M }    A_{nm} (\tau) |n \rangle {\rm e}^ {\sum_{\bf q} (f_{m{\bf q}}(\tau) b_{\bf q}^\dagger - {\rm {H.C.}})} | \bf 0 \rangle \times {\rm e}^ {\sum_{\tilde{\bf q}} (\tilde f_{m{\tilde{\bf q}}}(\tau) \tilde b_{\tilde{\bf q}}^\dagger - {\rm {H.C.}})} \tilde {| \bf 0 \rangle}
\end{equation}
where $|n \rangle$ numbers the hole states and $f_{m {\bf q}}$ ($\tilde f_{m{\tilde{\bf q}}}$) denote the displacement of the magnon mode with momentum $\bf q$ ($\tilde{\bf q}$) in the $m$th coherent state, and $|\bf 0 \rangle$ ($|\tilde { \bf 0} \rangle$) are the vacuum states for the ``physical" and ``tilde" magnon degrees of freedom.

The time-dependence of the variational parameters $A_{nm}(\tau)$, $f_{m {\bf q}}(\tau)$ and $\tilde f_{m{\tilde{\bf q}}}(\tau)$ is determined via the variational principle~\cite{zhao1, zhao2}
\begin{eqnarray}
\label{Euler}
\frac{d}{dt}\frac{\partial \mathcal{L}}{\partial \dot{\xi}_{j}^{\ast}}-\frac{\partial \mathcal{L}}{\partial \xi_{j}^{\ast}} = 0,
\end{eqnarray}
where the Lagrangian $\mathcal{L}$ is given by
\begin{eqnarray}
\label{Lagrangian}
\mathcal{L}&=&\frac{i}{2}\left[\langle {\rm D}_{2}^{M}(\tau)|\frac{\overrightarrow{\partial}}{\partial \tau}|{\rm D}_{2}^{M}(\tau)\rangle
-\langle {\rm D}_{2}^{M}(\tau)|\frac{\overleftarrow{\partial}}{\partial \tau}|{\rm D}_{2}^{M}(\tau)\rangle\right]-\langle{\rm D}_{2}^{M}(\tau)|{H_{\theta}}|{\rm D}_{2}^{M}(\tau)\rangle.
\end{eqnarray}

\section{\Large S3. Equations of motion for t-J model at finite temperature}
\renewcommand{\theequation}{S\arabic{equation}}
The Hamiltonian in the multi-$\rm D_2$ Ansatz is defined as
\begin{align}
\label{Hami}
L_{H_{\theta}} &=  \langle D_2^{\rm M} (\tau) | H_{\theta} | D_2^{\rm M} (\tau) \rangle =   \sum_{n} \sum_{j}^{M} \sum_{u}^{M} A_{nj}^{\ast} A_{nu} \sum_{\bf q} \omega_{\bf q} (f_{j \bf q}^{\ast} f_{u \bf q}-\tilde f_{j \bf q}^{\ast} \tilde f_{u \bf q}) R(f_j^{\ast}, f_u) \nonumber \\
& + \frac {tz} {\sqrt{N}} \sum_{\bf {kq}} \sum_{j}^{M} \sum_{u}^{M} A_{{\bf (k-q)}j}^ {\ast} A_{{\bf (k)}u} [{\rm cosh}(\theta_{\bf q})(u_{\bf q} \gamma_{\bf {k-q}}+v_{\bf q} \gamma_{\bf k}) f_{j{\bf q}}^{\ast}+{\rm sinh}(\theta_{\bf q})(u_{\bf q} \gamma_{\bf {k-q}}+v_{\bf q} \gamma_{\bf k}) \tilde f_{u{\bf q}}] R(f_j^{\ast}, f_u)\nonumber\\
& + \frac {tz} {\sqrt{N}} \sum_{\bf {kq}} \sum_{j}^{M} \sum_{u}^{M} A_{{\bf (k+q)}j}^ {\ast} A_{{\bf (k)}u} [{\rm cosh}(\theta_{\bf q})(u_{\bf {q}} \gamma_{\bf {k}}+v_{\bf {q}} \gamma_{\bf {k+q}}) f_{u{\bf q}}+{\rm sinh}(\theta_{\bf q})(u_{\bf {q}} \gamma_{\bf {k}}+v_{\bf {q}} \gamma_{\bf {k+q}}) \tilde f_{j{\bf q}}^{\ast}] R(f_j^{\ast}, f_u)\nonumber \\
\end{align}
where
$$R
(f_j^{\ast}, f_u) = {\rm {exp}} [\sum_{l} f_{jl}^{\ast} (\tau) f_{ul}(\tau)+\sum_{l} \tilde f_{jl}^{\ast} (\tau) \tilde f_{ul}(\tau)]
$$
is the Debye-Waller factor.

Thus the equation of motion for $A_{nu}$ assumes the form
\begin{align}
& i \sum_{n} \sum_{u}^{M} [\dot{A}_{nu} + A_{nu} \sum_{l} f_{jl}^{\ast} \dot{f}_{ul}+ A_{nu} \sum_{l} \tilde f_{jl}^{\ast} \tilde {\dot{f}}_{ul}] R(f_j^{\ast}, f_u) \nonumber\\
&= \sum_{u}^{M} A_{nu} \sum_{\bf q} \omega_{\bf q} (f_{j \bf q}^{\ast} f_{u \bf q}-\tilde f_{j \bf q}^{\ast} \tilde f_{u \bf q}) R(f_j^{\ast}, f_u) \nonumber\\
& + \frac {tz} {\sqrt{N}} \sum_{\bf {kq, k-q=n}} \sum_{u}^{M}  A_{{\bf (k)}u} [{\rm cosh}(\theta_{\bf q})(u_{\bf q} \gamma_{\bf {n}}+v_{\bf q} \gamma_{\bf k}) f_{j{\bf q}}^{\ast}+{\rm sinh}(\theta_{\bf q})(u_{\bf q} \gamma_{\bf {n}}+v_{\bf q} \gamma_{\bf k}) \tilde f_{u{\bf q}}] R(f_j^{\ast}, f_u)\nonumber\\
& + \frac {tz} {\sqrt{N}} \sum_{\bf {kq, k+q=n}} \sum_{u}^{M} A_{{\bf (k)}u} [{\rm cosh}(\theta_{\bf q})(u_{\bf {q}} \gamma_{\bf {k}}+v_{\bf {q}} \gamma_{\bf {n}}) f_{u{\bf q}}+{\rm sinh}(\theta_{\bf q})(u_{\bf {q}} \gamma_{\bf {k}}+v_{\bf {q}} \gamma_{\bf {n}}) \tilde f_{j{\bf q}}^{\ast}] R(f_j^{\ast}, f_u)
\end{align}

Similarly, the equations of motion for $f_{ul}$  and $\tilde f_{ul}$ are given by the formulas
\begin{align}
& i \sum_{n} \sum_{u}^{M} A_{nj}^{\ast} A_{nu} \dot{f}_{ul} R(f_j^{\ast}, f_u) + i \sum_{n} \sum_{u}^{M} [A_{nj}^{\ast} \dot{A}_{nu} + A_{nj}^{\ast} A_{nu}(\sum_{l} f_{jl}^{\ast} \dot{f}_{ul}+ \sum_{l} \tilde f_{jl}^{\ast} \tilde {\dot{f}}_{ul}) ] R(f_j^{\ast}, f_u) f_{ul} \nonumber \\
& =\sum_{n} \sum_{u}^{M} A_{nj}^{\ast} A_{nu} \sum_{\bf q} \omega_{\bf q} (f_{j \bf q}^{\ast} f_{u \bf q}-\tilde f_{j \bf q}^{\ast} \tilde f_{u \bf q}) R(f_j^{\ast}, f_u) f_{ul} + \sum_{m} \sum_{u}^{M} A_{nj}^{\ast} A_{nu} \omega_{\bf l} f_{u \bf l} R(f_j^{\ast}, f_u) \nonumber \\
& + \frac {tz} {\sqrt{N}} \sum_{\bf {kq}} \sum_{u}^{M} A_{{\bf (k-q)} j}^{\ast} A_{{\bf (k)}u} [{\rm cosh}(\theta_{\bf q})(u_{\bf q} \gamma_{\bf {k-q}}+v_{\bf q} \gamma_{\bf k}) f_{j{\bf q}}^{\ast}+{\rm sinh}(\theta_{\bf q})(u_{\bf q} \gamma_{\bf {k-q}}+v_{\bf q} \gamma_{\bf k}) \tilde f_{u{\bf q}}] R(f_j^{\ast}, f_u)f_{ul} \nonumber \\
& + \frac {tz} {\sqrt{N}} \sum_{\bf {kq}} \sum_{u}^{M} A_{{\bf (k-q)} j}^{\ast} A_{{\bf (k)}u}[{\rm cosh}(\theta_{\bf q})(u_{\bf {q}} \gamma_{\bf {k}}+v_{\bf {q}} \gamma_{\bf {k+q}}) f_{u{\bf q}}+{\rm sinh}(\theta_{\bf q})(u_{\bf {q}} \gamma_{\bf {k}}+v_{\bf {q}} \gamma_{\bf {k+q}}) \tilde f_{j{\bf q}}^{\ast}] R(f_j^{\ast}, f_u)f_{ul} \nonumber \\
& + \frac {tz} {\sqrt{N}} \sum_{\bf {k}} \sum_{u}^{M} A_{{\bf (k-l)} j}^{\ast} A_{{\bf (k)}u} [{\rm cosh}(\theta_{\bf l})(u_{\bf l} \gamma_{\bf {k-l}}+v_{\bf l} \gamma_{\bf k}) R(f_j^{\ast}, f_u)
\end{align}

and

\begin{align}
& i \sum_{n} \sum_{u}^{M} A_{nj}^{\ast} A_{nu} \tilde {\dot{f}}_{ul} R(f_j^{\ast}, f_u) + i \sum_{n} \sum_{u}^{M} [A_{nj}^{\ast} \dot{A}_{nu} + A_{nj}^{\ast} A_{nu}(\sum_{l} f_{jl}^{\ast} \dot{f}_{ul}+ \sum_{l} \tilde f_{jl}^{\ast} \tilde {\dot{f}}_{ul}) ] R(f_j^{\ast}, f_u) \tilde f_{ul}\nonumber\\
& =\sum_{n} \sum_{u}^{M} A_{nj}^{\ast} A_{nu} \sum_{\bf q} \omega_{\bf q} (f_{j \bf q}^{\ast} f_{u \bf q}-\tilde f_{j \bf q}^{\ast} \tilde f_{u \bf q}) R(f_j^{\ast}, f_u) \tilde f_{ul}- \sum_{n} \sum_{u}^{M} A_{nj}^{\ast} A_{nu} \omega_{\bf l} \tilde f_{u \bf l} R(f_j^{\ast}, f_u)\nonumber\\
& + \frac {tz} {\sqrt{N}} \sum_{\bf {kq}} \sum_{u}^{M} A_{{\bf (k-q)} j}^{\ast} A_{{\bf (k)}u} [{\rm cosh}(\theta_{\bf q})(u_{\bf q} \gamma_{\bf {k-q}}+v_{\bf q} \gamma_{\bf k}) f_{j{\bf q}}^{\ast}+{\rm sinh}(\theta_{\bf q})(u_{\bf q} \gamma_{\bf {k-q}}+v_{\bf q} \gamma_{\bf k}) \tilde f_{u{\bf q}}] R(f_j^{\ast}, f_u)\tilde f_{ul}\nonumber\\
& + \frac {tz} {\sqrt{N}} \sum_{\bf {kq}} \sum_{u}^{M} A_{{\bf (k+q)} j}^{\ast} A_{{\bf (k)}u} [{\rm cosh}(\theta_{\bf q})(u_{\bf {q}} \gamma_{\bf {k}}+v_{\bf {q}} \gamma_{\bf {k+q}}) f_{u{\bf q}}+{\rm sinh}(\theta_{\bf q})(u_{\bf {q}} \gamma_{\bf {k}}+v_{\bf {q}} \gamma_{\bf {k+q}}) \tilde f_{j{\bf q}}^{\ast}] R(f_j^{\ast}, f_u)\tilde f_{ul} \nonumber \\
& + \frac {tz} {\sqrt{N}} \sum_{\bf {k}} \sum_{u}^{M} A_{{\bf (k+l)}j}^{\ast} A_{{\bf k}u} [{\rm sinh}(\theta_{\bf l})(u_{\bf {l}} \gamma_{\bf {k}}+v_{\bf {l}} \gamma_{\bf {k+l}}) R(f_j^{\ast}, f_u)
\end{align}

The magnon population in  site $\bf d$ is expressed as
\begin{align}
N_{\bf d}(\tau) &= \langle D_2^M(\tau)|e^{iG} b_{\bf d}^{\dagger}b_{\bf d} e^{-iG}|D_2^M(\tau) \rangle \nonumber\\
&= \langle D_2^M(\tau)| \frac {1} {N} \sum_{\bf {q_1 q_2}} e^{-i(\bf {q_1-q_2}) {\bf d}} e^{iG} b_{\bf q_1}^{\dagger}b_{\bf q_2} e^{-iG}|D_2^M(\tau) \rangle \nonumber\\
&= \langle D_2^M(\tau)| \frac {1} {N} \sum_{\bf {q_1 q_2}} e^{-i(\bf {q_1-q_2}) {\bf d}} [{\rm cosh}(\theta_{\bf q1}){\rm cosh}(\theta_{\bf q2}) b_{\bf q_1}^{\dagger}b_{\bf q_2} + {\rm cosh}(\theta_{\bf q1}){\rm sinh}(\theta_{\bf q2}) b_{\bf q_1}^{\dagger} \tilde {b}_{\bf q_2}^{\dagger} \nonumber\\
& + {\rm sinh}(\theta_{\bf q1}){\rm cosh}(\theta_{\bf q2}) \tilde {b}_{\bf q_1} {b}_{\bf q_2}+ {\rm sinh}(\theta_{\bf q1}){\rm sinh}(\theta_{\bf q2}) \tilde {b}_{\bf q_1}  \tilde {b}_{\bf q_2}^{\dagger}] |D_2^M(\tau) \rangle \nonumber\\
& =\frac {1} {N}  \sum_{\bf {q_1 q_2}} e^{-i(\bf {q_1-q_2}) {\bf d}} \sum_n \sum_{j,u}^M A_{nj}^{\ast} A_{nu} [f_{j (\bf q_1)}^{\ast}(\tau) f_{u (\bf q_2)}(\tau){\rm cosh}(\theta_{\bf q1}){\rm cosh}(\theta_{\bf q2}) \nonumber\\
& + f_{j (\bf q_1)}^{\ast}(\tau) \tilde {f}_{j (\bf q_2)}^{\ast}(\tau){\rm cosh}(\theta_{\bf q1}){\rm sinh}(\theta_{\bf q2}) + \tilde {f}_{u (\bf q_1)}(\tau) {f}_{u (\bf q_2)}(\tau){\rm sinh}(\theta_{\bf q1}){\rm cosh}(\theta_{\bf q2})\nonumber\\
&+ \tilde {f}_{j (\bf q_2)}^{\ast}(\tau) \tilde {f}_{u (\bf q_1)}(\tau){\rm sinh}(\theta_{\bf q1}){\rm sinh}(\theta_{\bf q2})]R(f_j^{\ast},f_u) + \frac {1} {N}  \sum_{\bf {q_1}}  \sum_n \sum_{j,u}^M A_{nj}^{\ast} A_{nu} {\rm sinh}^{2} (\theta_{\bf  q1})R(f_j^{\ast},f_u)
\end{align}

\end{document}